%% file: main.tex
\documentclass[a4paper, amsfonts, amssymb, amsmath, reprint, showkeys, nofootinbib, twoside,superscriptaddress]{revtex4-1}
\usepackage[english]{babel}
\usepackage[utf8]{inputenc}
\usepackage[colorinlistoftodos, color=green!40, prependcaption]{todonotes}
\input{preamble}
\usepackage{amsfonts}
\usepackage{stmaryrd}
\usepackage[bbgreekl]{mathbbol}

\DeclareSymbolFontAlphabet{\mathbbm}{bbold}
\DeclareSymbolFontAlphabet{\mathbb}{AMSb}%

\usepackage{caption}
\usepackage{bm}
\usepackage{adjustbox}

\usepackage{ dsfont }
\usepackage{multirow}
\usepackage{tabularx,ragged2e,booktabs}
\usepackage{hhline}
\usepackage{cases}
\usepackage{subcaption}
\usepackage{calrsfs}
\usepackage{verbatim}
\DeclareMathAlphabet{\pazocal}{OMS}{zplm}{m}{n}

\newcommand{\Cb}{\pazocal{C}}
\newcommand{\Db}{\pazocal{D}}

\usepackage[pdftex, pdftitle={Article}, pdfauthor={Author}]{hyperref} 

\begin{document}

\title{Multivariate Quadratic Hawkes Processes -- Part I: Theoretical Analysis}

\author{Cécilia Aubrun}
\affiliation{Chair of Econophysics and Complex Systems, \'Ecole polytechnique, 91128 Palaiseau Cedex, France}
\affiliation{LadHyX UMR CNRS 7646, \'Ecole polytechnique, 91128 Palaiseau Cedex, France}

\author{Michael Benzaquen}
\email{michael.benzaquen@polytechnique.edu}
\affiliation{Chair of Econophysics and Complex Systems, \'Ecole polytechnique, 91128 Palaiseau Cedex, France}
\affiliation{LadHyX UMR CNRS 7646, \'Ecole polytechnique, 91128 Palaiseau Cedex, France}
\affiliation{Capital Fund Management, 23 Rue de l’Universit\'e, 75007 Paris, France}
\author{Jean-Philippe Bouchaud}
\affiliation{Chair of Econophysics and Complex Systems, \'Ecole polytechnique, 91128 Palaiseau Cedex, France}
\affiliation{Capital Fund Management, 23 Rue de l’Universit\'e, 75007 Paris, France}
\affiliation{Académie des Sciences, 23 Quai de Conti, 75006 Paris, France\smallskip}

\date{\today} 

\begin{abstract}
Quadratic Hawkes (QHawkes) processes have proved effective at reproducing the statistics of price changes, capturing many of the stylised facts of financial markets. Motivated by the recently reported strong occurrence of endogenous co-jumps (simultaneous price jumps of several assets) we extend QHawkes to a multivariate framework (MQHawkes), that is considering several financial assets and their interactions. Assuming that quadratic kernels write as the sum of a time-diagonal component and a rank one (trend) contribution, we investigate endogeneity ratios and the resulting stationarity conditions. We then derive the so-called Yule-Walker equations relating covariances and feedback kernels, which are essential to calibrate the MQHawkes process on empirical data. Finally, we investigate the volatility distribution of the process and find that, as in the univariate case, its tail exhibits power-law behavior, with a unique exponent that can be exactly computed in some limiting cases.  
\end{abstract}


\maketitle
\section*{Introduction} \label{sec:intro}


Modelling the volatility of financial assets is a significant challenge for academics, market participants and regulators alike. In fact, models describing the statistics of price changes are widely used for, e.g., risk control and derivative pricing. When not in line with the behaviour of real markets, these models can lead to disappointing outcomes, or even major mishaps. Thus, it is important to account for the main stylized facts observed in real price time series, in particular: fat tails of the returns distribution, multi-timescale volatility clustering, price-volatility correlations and time reversal asymmetry. 

Among the most widely used models, one can quote stochastic volatility models and GARCH models. However, these models often capture only poorly the above stylized facts.  Indeed, the most popular stochastic volatility models lead to thin tails in the returns distribution, and to strict time reversal symmetry.  Whereas GARCH models do present time reversal asymmetry, such asymmetry is actually stronger than that observed in real markets. Furthermore, GARCH volatility clustering is governed by a single time scale. Some of these problems are elegantly solved by the family of rough volatility models \cite{gatheral2018volatility}, in particular the multifractal random walk which is but  a special member of that family \cite{bacry2001multifractal,wu2022rough}.   

Hawkes processes, on the other hand, provide an interesting alternative to more traditional models, while clearly highlighting the feedback loop at the origin of the stylized facts characterizing financial prices. However, Jaisson and Rosenbaum~\cite{jaisson2015limit} have shown that the continuous time limit of a (near-critical) Hawkes process is a fractional CIR Heston process, which is characterized by tails that are too thin to reproduce empirical returns distribution. Furthermore, such a process is again time reversal invariant.

\begin{figure}
    \centering
    \includegraphics[width=\linewidth]{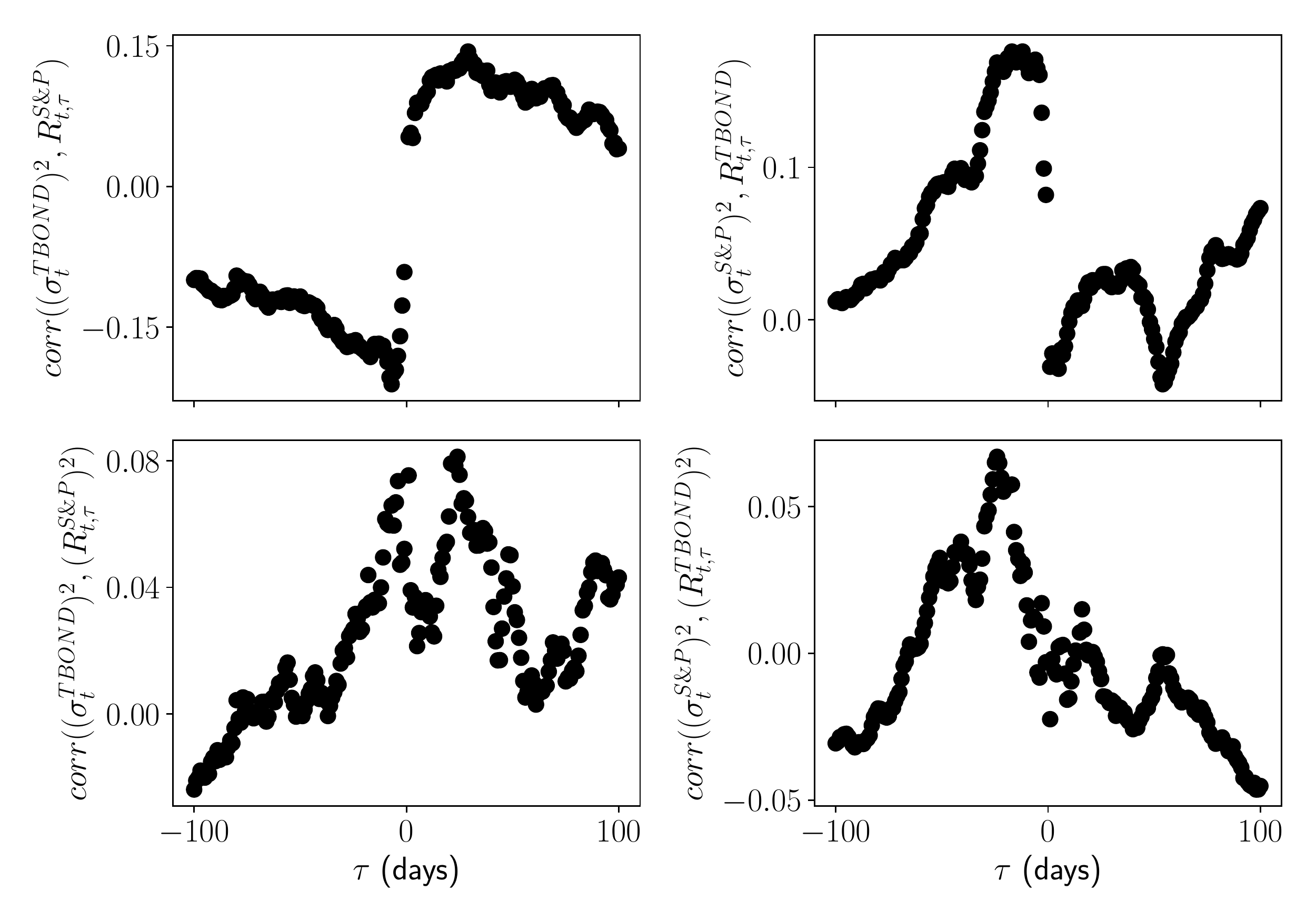}
    \caption{{Top: Cross asset leverage effect between the S\&P500 futures and the US 10 years Treasury bond futures (``{\sc tbond}''). We plot the correlation between the high-frequency daily volatility $\sigma_t^2$ and the past and future returns $R_{t,\tau}$ on scale $\tau$. Bottom: Cross-Zumbach effect between the {S\&P500} and the {\sc tbond} future contracts (after factoring in the dominant leverage effect). We plot the correlation between the volatility $\sigma_t^2$ and the squared returns $R_{t,\tau}^2$ on scale $\tau$, showing a clear asymmetry between past and future, specially from past {\sc tbond} trends towards S\&P500 volatility.  The daily volatility $\sigma^2$ is computed as the mean of the square returns on 5 min windows over one day (overnight excluded). Aggregated returns are defined as $R_{t,\tau} := P^{\text{close}}_{t-1}/P^{\text{open}}_{t-1+\tau}-1$ for $\tau<0$ and $R_{t,\tau} := P^{\text{close}}_{t+1+\tau}/P^{\text{open}}_{t+1}-1$ for $\tau>0$. The lag $\tau$ is in days. The data covers the period 2013-2018.}
    }
    \label{fig:stylisedFact}
\end{figure}

To alleviate these problems, Donier, Blanc, Bouchaud in \cite{blanc2017quadratic} introduced the family of ``Quadratic'' Hawkes (QHawkes) processes, which encodes a feedback between past price {\it trends} and future volatility, a clear empirical effect first discovered by Zumbach \cite{zumbach2009time}, see also \cite{chicheportiche2014fine}. The quadratic feedback allows one to overcome the limitations of standard (linear) Hawkes processes. QHawkes processes naturally generate fat tail distributions and time reversal asymmetry even in the continuous time limit.

\begin{figure}[h]
    \centering
    \includegraphics[width=\linewidth]{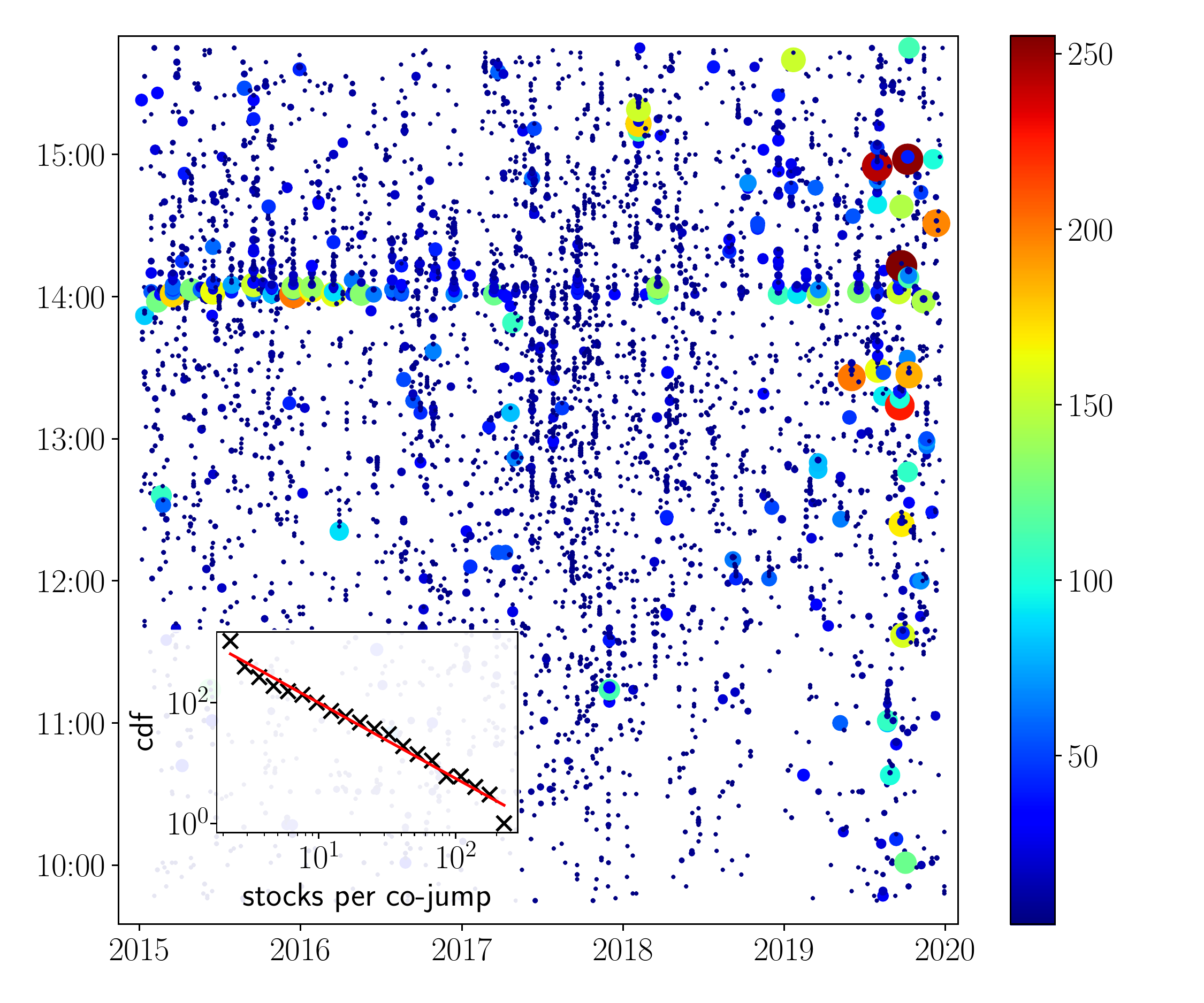}
    \caption{Time series of 1 minute bins co-jumps in a set of 295 US stocks over 5 years. The
horizontal axis corresponds to the day in the sample and the vertical axis gives the time of day. The size and the color of the circle
encodes the number of stocks simultaneously jumping in a given minute (see color bar). The insert shows the cumulative distribution function of the number of stocks in co-jumps for jumps that were classified as endogenous according to the methodology developed in \cite{marcaccioli2022exogenous}, the slope is $-1.25$.}
    \label{fig:cojump2}
\end{figure}

In this paper, we generalize the monovariate QHawkes process of Ref. \cite{blanc2017quadratic} to the multivariate case, allowing one to capture how the trend on one asset can impact the volatility of another asset. {Like for the univariate case, we expect the multivariate expansion to capture some stylised facts that are not accounted for within a linear Hawkes framework. In order to motivate our study, we illustrate in Figure \ref{fig:stylisedFact} the existence of a {\it cross-asset} leverage and Zumbach effects, i.e. the fact that past trends on the S\&P500 (resp. {\sc tbond}) increases volatility of the {\sc tbond} (resp. S\&P500), in a way that is asymmetric between past and future, much as the usual ``self'' Zumbach effect. Figure \ref{fig:cojump2} reveals high activity episodes across the whole stock market, with the number of simultaneous co-jumps being distributed as a power-law. This suggests the existence of a propagation phenomenon similar to a branching process, which requires a specific non-linear cross-asset interaction of the type considered in this paper.

The outline of the paper is as follows. In Section~\ref{sec:develop}, we define the multivariate QHawkes model in full generality. Section~\ref{sec:cov} establishes generalized Yule-Walker equations that relate 2- and 3-point correlation functions to the various feedback kernels appearing in the definition of the model. These relations are the starting point for the calibration of the model to empirical data, a topic we will explore in a follow-up, companion paper \cite{ustocome}. Section~\ref{sec:vol} investigates the volatility distribution generated by the model and establishes the power-law nature of the tails, with an exponent that we analytically compute in several special cases. In Section~\ref{sec:conclusion} we conclude.}\newline


Besides, a notation table is provided after the bibliography to ease the reading (See Appendix \ref{app:notations}).

\section{Model Presentation} \label{sec:develop}

In this  section, we introduce  Multivariate Quadratic Hawkes processes (MQHawkes). We start by recalling the definition of Hawkes and Quadratic Hawkes processes in their univariate form. We then define their extension to the multivariate case, with a special focus on the bivariate case. Next we introduce a much simplified specification, where the quadratic feedback can be factorized into a product of \textit{trend indicators} -- a limit called ZHawkes in \cite{blanc2017quadratic}, after Zumbach's seminal work on the time reversal asymmetry of financial time series.  We discuss endogeneity ratios and stationarity conditions for MQHawkes within the ZHawkes framework. Finally, we further generalize the MQHawkes model to take into account the possible intrinsic correlations between the underlying (inhomogeneous) Poisson processes.

\subsection{Hawkes and QHawkes Processes}

Hawkes processes were first introduced to model earthquakes \cite{koshelev1981testing,zhuang2012basic}. Their ability to consider the past events in the probability of a future event was indeed very useful to reproduce the aftershock phenomenon: an earthquake is in fact more likely to occur after another just took place. This same property appeared to  hold in a variety of fields, like biological neural networks \cite{osorio2010epileptic,sornette2010prediction}, financial markets \cite{bacry2015hawkes,fosset2021non}, and also crime rates or riot propagation \cite{mohler2011self,bonnasse2018epidemiological}.

A Hawkes process $(N_t)_{t\geq0}$ is an inhomogenous\footnote{Here, inhomogenous means that the intensity of the process is time dependent.} Poisson process with intensity $\lambda_t$ defined from the past realisations of the process itself, according to:\newline
\begin{equation}\label{eq:Hawkes_def}
    \lambda_t= \lambda_\infty + \int_{-\infty}^t\phi(t-u)\, {\rm d}N_u
\end{equation}
    with
\begin{equation}
    n = \int_\mathbb{R}\phi(u)\, {\rm d}u < 1,
\end{equation}
where $\lambda_\infty$ is the baseline intensity, $\phi$ is the feedback kernel and $n$ represents the ``endogeneity ratio'', which needs to be strictly less than unity in order to guarantee that the process reaches a stationary state.

Like Hawkes processes, QHawkes processes are inhomogenous Poisson processes with an intensity defined from past events. 
In the context of financial markets, the event process $(N_t)_{t>0}$ represents the sequence of prices changing events, where the price change is defined by   ${\rm d}P_t = \epsilon_t \psi {\rm d}N_t$~\cite{blanc2017quadratic} where $\epsilon_t = \pm 1$ (random sign), and $\psi$ is the tick size (the smallest unit of price change) which we set to one henceforth. The intensity self-exciting feedback loop  is now defined on the returns ${\rm d}P_t$ rather than on the events ${\rm d}N_t$. Hence, the sign of the returns influences the process intensity \cite{blanc2017quadratic}: 
\begin{equation}\label{eq:Lambda_def_QH}
\begin{aligned}
    \lambda_t =& \lambda_{\infty} +  \int^t_{-\infty}L(t-s)\,{\rm d}P_s\\
    &+\iint^t_{-\infty} K(t-s,t-u)\, {\rm d}P_s{\rm d}P_u,
\end{aligned}
\end{equation}
where $\lambda_\infty$ is again the baseline intensity, $L(\cdot)$ is the so-called leverage kernel (capturing the increase of activity following a drop in prices) and $K(\cdot,\cdot)$ is the quadratic kernel. 

Following {empirical findings of} \cite{blanc2017quadratic}, we simplify the quadratic kernel $K$ assuming it can be written as a general time-diagonal and a rank one contribution:
\begin{equation}
    K(s,u) = \phi(s) \delta(s-u) + k(s) k(u).
\end{equation}
This defines what we shall refer to as the ZHawkes model:
\begin{equation}\label{eq:Lambda_def_QZ}
\begin{aligned}
    \lambda_t =& \lambda_{\infty} +  \int^t_{-\infty}L(t-s)\,{\rm d}P_s +  \int^t_{-\infty}\phi(t-s)\,{\rm d}N_s\\
    &+\left( \int^t_{-\infty} k(t-s)\, {\rm d}P_s \right)^2,
\end{aligned}
\end{equation}
where we have used $({\rm d}P_s)^2 =  {\rm d}N_s$. Note that the diagonal term precisely reproduces the standard Hawkes feedback. 

\subsection{MQHawkes Processes}

Let us now consider $N$ financial assets with prices $(P^i_t)_{t\geq0,i=1, \dots,N}$. We associate such price processes with jump processes $(N^i_t)_{t\geq0}$, with: 
\begin{equation}
    {\rm d}P^i_t = \epsilon_i \psi_i {\rm d}N^i_t, 
\end{equation} 
where we set henceforth $\psi_i \equiv 1$, $\forall i$, without loss of generality. Each jump process $(N^i_t)_{t\geq0}$ is a conditionally independent\footnote{This means that for given intensities $\lambda_{i,t}$, the inhomogeneous Poisson processes ${\rm d}N^i_t$ are independent. See further down for the case of correlated jump processes.} Quadratic Hawkes process with intensity $(\lambda_{i,t})_{t\geq0}$  
\begin{equation}\label{eq:Lambda_def_MQH}
\begin{aligned}
    \lambda_{i,t} =& \lambda_{i,\infty} + \sum_{j=1}^N  \int^t_{-\infty}L_j^i(t-s)\,{{\rm d}P^j_s} \\
    &+\sum_{j \leq k}^N  \iint^t_{-\infty} K_{jk}^i(t-s,t-u)\, {{\rm d}P^j_s}\, {\rm d}P^k_u.
\end{aligned}
\end{equation}
Note that the superscript in the kernels indicates which asset is affected by the feedback, whereas subscripts indicate which assets are responsible for it.


In the following we shall generically call $\mathbb{K_{\text{d}}}$ the ``diagonal'' feedback terms, i.e. the 
kernels $K_{jk}^i$ with $j=k$. These terms describe the quadratic feedback from two price changes of the same asset $j$ onto the activity of asset $i$. Similarly, $\boldsymbol{K_{\times}}$  describes cross effects, i.e. $K_{jk}^i$ with $j < k$, from two different assets $j \neq k$ onto the activity of asset $i$. In order to guarantee that intensities $\lambda_t^i$ remain positive at all times, kernels ${L}, \mathbb{K}_{\text{d}}$ and $\mathbf{K}_{\times}$ need to verify some conditions, which are detailed in Appendix \ref{app:positiveLambda}.  

Although all these terms could in principle play a role,  in the present paper we shall restrict to cross terms $K_{jk}^i$ such that either $j$ or $k$ are equal to $i$. The intuitive meaning of such terms will become clear below in the context of ZHawkes processes; in particular we shall see why terms with $j \neq i$ and  $k \neq i$ are not expected to play a large role in practice. 

For the sake of clarity we will mainly focus on the bivariate case $N=2$, for which the leverage kernel and the diagonal quadratic kernel are $2 \times 2$ matrices:
\begin{equation*}
    \mathbb{L} := \begin{pmatrix}
    L^1_1&L^1_2\\L^2_1&L^2_2
    \end{pmatrix}; \quad \mathbb{K_{\text{d}}} := \begin{pmatrix}
    K^1_{11}&K^1_{22}\\K^2_{11}&K^2_{22}
    \end{pmatrix},
\end{equation*}
whereas the cross quadratic kernel is a  vector
\begin{equation*}
    \boldsymbol{K_{\times}} := \begin{pmatrix}
    K^1_{12}\\K^2_{12}
    \end{pmatrix}.
\end{equation*}
In the following $2 \times 2$ matrices and $2$D vectors are respectively noted $\mathbb{A}$ and 
$\boldsymbol{A}$, for example $
    \boldsymbol{\lambda} := 
   ( \lambda_1, \lambda_2)$.
We will also need to distinguish the ``time diagonal'' of a matrix $\mathbb{A}(\tau_1,\tau_2)$, by which we mean $\mathbb{A}(\tau,\tau)$, from the ``diagonal'' of $\mathbb{A}$, which refers to the diagonal components in asset space $\mathbb{A}_{ii}$.

\subsection{ZHawkes Model}

The multivariate generalisation of the ZHawkes model amounts to assuming that the quadratic kernels $\mathbb{K_{\text{d}}}$ and $\boldsymbol{K_{\times}}$ can be written as arbitrary singular time-diagonal and  time-rank-one contributions, to wit:
\begin{align*}
    K_{jj}^i(s,u) &= ({\phi_{\text{d}}})_{jj}^i(s)\, \delta(s-u)+ k^i_{jj}(s) k^i_{jj}(u)\\
    K_{ij}^i(s,u) &= ({\phi_{\times}})_{ij}^i(s)\,\delta(s-u)+k^i_{ji}(s) k^i_{ij}(u) \quad (i < j).
\end{align*}
However, as long as we consider independent Poisson processes, we can set  $\boldsymbol{\phi_{\times}}=\boldsymbol{0}$. This is because the two processes will almost never jump simultaneously, such that for all $u$,  ${\rm d}P^i_u{\rm d}P^j_u=0$ when $i \neq j$ (see however Eq.~ \eqref{eq:def_Lambda_dep} below when  co-jumps are taken into account). Actually in the so-called Thinning Algorithm \cite{ogata1981lewis} for multivariate processes, commonly used to simulate inhomogenous Poisson process, at most one process can be jumping at each time step.

\subsection{Endogeneity Ratio and Stationarity Condition}

{
 The endogeneity ratios indicate by how much, on average, the feedback loop  contributes to the future of the process, and thus, whether the process is stationary or not.} To define them for the MQHawkes we use analogies with  univariate QHawkes and  multivariate linear Hawkes. 

\subsubsection{Mean Intensity}

Since price changes are centred and processes are assumed to be independent, we have $\mathbb{E}({\rm d}\boldsymbol{P})=0$ and $\mathbb{E}({\rm d}P^i_s {\rm d}P^j_u)= \delta_{ij}\delta_{s,u}  \Bar{\lambda}^i {\rm d}s$. Using this, we find that the vector of mean intensities  $\boldsymbol{\Bar{\lambda}}${, when finite,} writes:
\begin{equation}\label{eq:stat_ind}
\begin{aligned}
        \boldsymbol{\Bar{\lambda}} &= \left(\mathbb{I}-\int_{0}^\infty \mathbb{K}_{\text{d}}(s,s) {\rm d}s\right)^{-1}\boldsymbol{\lambda_\infty}.
\end{aligned}
\end{equation}

This expression must be interpreted with care when $\mathbb{K}_{\text{d}}(s,u)$ contains a singular time-diagonal contribution ${\bbphi_{\text{d}}}(s)\, \delta(s-u)$. In such a case, and throughout this paper, we will interpret $\mathbb{K}_{\text{d}}(s,s)$ as $\mathbb{K}_{\text{d,reg.}}(s,s) + \bbphi_{\text{d}}(s)$, where $\mathbb{K}_{\text{d,reg.}}$ is the regular part of the diagonal quadratic feedback.
Equation~\eqref{eq:stat_ind}  shows that the mean intensity diverges when the spectral radius\footnote{The spectral radius of a square matrix is the maximum of the absolute values of its eigenvalues.} of the matrix $\int_{0}^\infty \mathbb{K}_{\text{d}}(s,s) {\rm d}s $ tends to one from below.

\subsubsection{Endogeneity Ratio}\label{sec:endo}

In the multidimensional case, the endogeneity ratio of a standard linear Hawkes process is defined by the spectral radius of the kernel matrix $\bbphi$ involved in the expression of $\boldsymbol{\Bar{\lambda}}$ \cite{bacry2015hawkes}. More precisely, introducing 
\begin{align*}
    \lVert f \rVert := \int_0^{+\infty}f(s){\rm d}s
\end{align*} 
one constructs the matrix $\lVert (\phi)_j^i \rVert$ and determines its eigenvalue with largest modulus, which in turn defines the dominant endogeneity ratio $n$.  Thus, $n=\rho_{\text{spectral}}(\lVert \bbphi \rVert)$, where $\rho_{\text{spectral}}$ is the spectral radius and $\lVert \bbphi \rVert$ stands for the matrix $\lVert (\phi)_j^i \rVert$. 

For a general univariate QHawkes model, one can always decompose the endogeneity ratio $n$ as the sum of the Hawkes endogeneity ratio $n_H$ associated with the singular time-diagonal contribution to $K(\cdot,\cdot)$, and a regular contribution $n_Q$: 
\begin{equation}
    n = \int_0^{+\infty} \phi(s){\rm d}s  + \int_0^{+\infty}K_{\text{reg.}}(s,s){\rm d}s := n_H + n_Q.
\end{equation}
In the special case of a ZHawkes quadratic kernel, the regular contribution 
reads:
\begin{equation}
    n_Q = n_Z = \int_0^{+\infty} k^2(s){\rm d}s.
\end{equation}

For a general MQHawkes, the endogeneity ratio is then defined by the spectral radius of $\int_{0}^\infty \mathbb{K}_{\text{d}}(s,s) {\rm d}s$, $n = \rho_{\text{spectral}}(\lVert \mathbb{K}_{\text{d}} \rVert)$.
However, when decomposing the kernel as a singular time-diagonal contribution and a regular contribution, one must be careful about the fact that the two matrices 
$\mathbb{K}_{\text{d,reg.}}(s,s)$ and $\bbphi_{\text{d}}$ do {\it not} commute in general. Hence the spectral radius defining the endogeneity ratio cannot be simply expressed as the sum of a Hawkes contribution $n_H$ (i.e. the spectral radius of $\lVert \bbphi_{\text{d}}  \rVert$) and of a regular or Zumbach contribution (i.e. the spectral radius of $\lVert \mathbb{K}_{\text{d,reg.}} \rVert$).

\subsubsection{Stationarity Condition}

For linear Hawkes processes to be stationary, the endogeneity ratio $n$ needs to be strictly less than one. Hence, only processes with finite mean intensity $\boldsymbol{\Bar{\lambda}}$ can be stationary. In the presence of a quadratic feedback, the situation is more intricate. In a previous communication~\cite{aubrun2022hawkes} we found that for a univariate Z-Hawkes process to be stationary one needs $n_H<1$, but not necessarily $n = n_H + n_Q < 1$. When $n_H < 1$ and $n > 1$ one has a stationary process with an infinite mean. In the present multivariate framework, we conjecture that a similar situation holds, with the following definitions: let $n_H$ be the spectral radius of $\lVert  \bbphi_{\text{d}} \rVert$, {and  $n=\rho_{\text{spectral}}(\lVert \mathbb{K}_{\text{d}} \rVert)$, with in general $n\neq n_H+n_Q$.} Then, there exists a value $n^\star$ such that:
\begin{itemize}
    \item if $n_H<1$ and $n < n^\star$, the process is stationary with finite mean intensity;
    \item if $n_H < 1$ and $n > n^\star$ the process is stationary with infinite mean intensity.
    \item if $n_H > 1$, the process explodes and no stationary state can be reached. 
\end{itemize}
We have performed numerical simulations (not shown) that support this conjecture. The value of the critical point $n^\star$ is however difficult to compute in the general case, but in the specific case of a weakly anisotropic two dimensional ZHawkes model, some progress can be made, see Section \ref{sec:anisotropic} and Eq. \eqref{eq:n-star}. In particular, in the isotropic case, the univariate result $n^\star=1$ is recovered.
 

\subsection{Correlated Poisson Processes}

One may wonder if the hypothesis of independence of the ${\rm d}N_t^i$ for different ${i = 1, \dots, N}$ is not too strong to faithfully account for events happening in financial markets.  In fact, as in \cite{bormetti2015modelling}, we find that {\it co-jumps} (i.e. simultaneous jumps in the price of different assets within 1 minute bins) occur fairly frequently, adding a new stylised fact to consider in this multivariate framework. In order to investigate co-jumps empirically, we  use a jump detection method (see \cite{marcaccioli2022exogenous}) on 295 large cap. US stock prices from January 2015 to December 2020. We then count how many stocks display anomalous price jumps in a given minute, and represented such counts in Figure \ref{fig:cojump2}. Co-jumps are clearly seen to occur. On average, there are 4.5 co-jumps per day, and up to 68 co-jumps in one day. Using \cite{marcaccioli2022exogenous} to classify each jump as endogenous or exogenous, we compute the cumulative density function of number of stocks included in endogenous co-jumps which displays a power law of slope $-1.25$ (see insert in Fig.~\ref{fig:cojump2}).

Co-jumps may be due to either a common exogenous shock (like an external piece of news affecting several stocks), or to some endogenous instability triggering a jump for one given stock, which propagates to other stocks. The very interesting question of the exogenous/endogenous nature of co-jumps clearly needs a more refined investigation, in the spirit of  \cite{marcaccioli2022exogenous},  and is left for future work.

In \cite{bormetti2015modelling}, the authors show that multivariate linear Hawkes models with independent realisations of the Poisson process do not satisfactorily reproduce co-jumps. Here we propose a way to enforce  correlations between Poisson processes, and  allow one to account for co-jumps within bivariate QHawkes processes.

\subsubsection{Bivariate Poisson Processes}

We focus on the bivariate case $N=2$. The extension to $N > 2$ is also possible, although beyond the scope of the present paper. In order to allow for the possibility of co-jumps, i.e. such that ${\rm d}P^1_t{\rm d}P^2_t \neq 0$, we consider three independent QHawkes counting processes $(N^{1}_t,N^{2}_t,N^{\rm{c}}_t)_{t\geq0}$ with intensities $(\mu_{1,t},\mu_{2,t},\mu_{{\rm{c}},t})$ defined from past returns: 
\begin{equation}\label{eq:def_Lambda_dep}
\begin{aligned}
      \mu_{a,t} =& \mu_{a,\infty} + \sum_{j\in\{1,2\}} \int^t_{-\infty} L^a_{j}(t-s){\rm d}P^j_s\\&+\sum_{j \leq k\in\{1,2\}}\iint^t_{-\infty} Q^a_{jk}(t-s,t-u){\rm d}P^j_s{\rm d}P^k_u,
\end{aligned}
\end{equation}
with $a=1,2, \rm{c}$, and we model price moves as: 
\begin{equation}
\begin{aligned}
      {\rm d}P^1_t &= \epsilon_t^1 ({\rm d}N^1_t+{\rm d}N_t^{\rm{c}})\\
    {\rm d}P^2_t &= \epsilon_t^2  ({\rm d}N^2_t+{\rm d}N_t^{\rm{c}}),
\end{aligned}
\end{equation}
where $\epsilon^i_t = \pm 1$, with $\mathbb{E}(\epsilon^i_t)=0$ and $\mathbb{E}(\epsilon^1_t \epsilon^2_t)=\rho_t$. While the sign correlation $\rho_t$ could indeed be time dependent, here we will assume that $\rho_t = \rho$ is independent of time. Thus, price moves have both an idiosyncratic part, represented by ${\rm d}N^i$, $i\in\{1,2\}$, and a common part ${\rm d}N^{\rm{c}}$, which make co-jumps possible\footnote{Note that if we relax the assumption that ${\rm d}P$ can only be equal to one tick, one can build a more general model \begin{align*}
    {\rm d}P^1_t = \epsilon^1_t
({\rm d}N^1_t + C_1{\rm d}N^c_t ) \\
{\rm d}P^2_t = \epsilon^2_t({\rm d}N^1_t + C_2{\rm d}N^c_t ) 
\end{align*} which would lead to a richer  correlation structure. Such a model would mean that co-jumps have a different size than usual price changes.} (hence the subscript $\rm{c}$). In fact, {because ${\rm d}N_t^{\rm{c}}\in \{0,1\}$, then $({\rm d}N_t^{\rm{c}})^2={\rm d}N_t^{\rm{c}}$,} one now has:
\begin{equation}
    {\rm d}P_t^1 {\rm d}P_t^2 =  \epsilon^1_t \epsilon^2_t {\rm d}N_t^{\rm{c}}.
\end{equation}
We can then define the total intensities $(\lambda_{1,t},\lambda_{2,t})_{t\geq0}$ of the price jumps $(P^1_t,P^2_t)_{t\geq0}$, according to the definition of Poisson processes: 
\begin{equation}
    \begin{cases}
    \lambda_{1,t} := \mu_{1,t} + \mu_{{\rm{c}},t}\\
    \lambda_{2,t} := \mu_{2,t} + \mu_{{\rm{c}},t},
    \end{cases}
\end{equation}
and notice that the structure of the equations governing the dynamics of intensities is conveniently the same as in the independent case, with the following identifications:  
\begin{align*}
    L^i_1 &\xleftarrow[]{} L^i_{1}+L^{\rm{c}}_{1} &
    L^i_2 &\xleftarrow[]{} L^i_{2}+L^{\rm{c}}_{2}\\
    K^i_{11} &\xleftarrow[]{} Q^i_{11}+Q^{\rm{c}}_{11}&
    K^i_{22} &\xleftarrow[]{} Q^i_{22}+Q^{\rm{c}}_{22}\\
    K^i_{12} &\xleftarrow[]{} Q^i_{12}+Q^{\rm{c}}_{12} & 
    \lambda^i_\infty  &\xleftarrow[]{} \mu^i_\infty + \mu^{\rm{c}}_\infty.
\end{align*}

\subsubsection{Mean Intensity}

Now correlations between instantanous price changes are non-zero, i.e.  $\mathbb{E}({\rm d}P^1_s {\rm d}P^2_s) = \rho \overline{\mu^{\rm{c}}} {\rm d}s\neq0$ (see Appendix \ref{app:stationaryConditionBiv}), the expression for the mean intensity $\boldsymbol{\Bar{\lambda}}$ is modified and reads, for regular kernels:
\begin{align}
    \boldsymbol{\Bar{\lambda}} =& \Big(\mathbb{I} - \int_{0}^\infty \mathbb{K}_{\text{d}}(s,s){\rm d}s\Big)^{-1}\\ \nonumber &\Big(\boldsymbol{\lambda_\infty} + \rho\overline{\mu^{\rm{c}}}
    \int_{0}^\infty \mathbf{K_{\times}}(s,s){\rm d}s\Big),
\end{align}
with:
\begin{equation*}
   \overline{\mu^{\rm{c}}} =  \frac{\mu_\infty^{\rm{c}} +  \boldsymbol{\Bar{\lambda}}^\top \cdot \int_{0}^\infty \boldsymbol{Q}^{\rm{c}}(s,s){\rm d}s }{1 - \rho \int_{0}^\infty Q_{12}^{\rm{c}}(s,s){\rm d}s},\
\end{equation*}
and 
\[
\boldsymbol{Q}^{\rm{c}}=\begin{pmatrix}{Q}^{\rm{c}}_{11}\\{Q}^{\rm{c}}_{22}\end{pmatrix} .
\]
 In the presence of a singular contribution to the $K$'s, one should again take it into account by substituting all ${K}(s,s)$ by the corresponding ${K}_{\text{reg}}(s,s) + {\phi}(s)$, and similarly for $Q$'s. The stationarity conditions are now that:
\begin{enumerate}
    \item The spectral radius of the Hawkes kernel $||\bbphi_{\text{d}}||$ is strictly less than 1
    \item The time-diagonal of the co-jump kernel must be such that: 
    \[ 
    \left\vert \rho \int_{0}^\infty Q_{12}^{\rm{c}}(s,s){\rm d}s \right\vert < 1.
    \]
\end{enumerate}

\section{Covariance Structure \& Yule-Walker Equations} \label{sec:cov}
    
Of course, the feedback kernels $L$ and $K$ cannot be directly observed in data. However, as we now show, they can be computed from covariance functions, which can easily be estimated from empirical data. Here we introduce the covariance structures of a multivariate QHawkes process, thereby establishing the matrix Yule-Walker equations (which link covariance structures and QHawkes kernels). 

In order to fully characterise the kernels $\mathbb{K}_{\text{d}}$ and $\mathbf{K}_{\cross}$, we need covariance structures containing information on both the diagonal part of the kernels ($\tau_1 = \tau_2)$ and their non diagonal part $\tau_1\neq \tau_2$. When time is discretized on a grid up to lag $q$, the number of unknowns is, for $N$ assets, $(q+{q(q-1)}/{2})\times N^2$ for $\mathbb{K}_\text{d}$ and  ${q(q-1)}\times N(N-1)$ unknowns for $\mathbf{K}_{\cross}$ (with $i=j$ or $i=k$ as considered in this paper and without explicit co-jumps).

Now, Equation \eqref{eq:C_YuleWalker_mQH} below on two-point correlations can only provide $q\times N(N+1)/2$ equations; three-point correlations are thus also needed to fully determine these kernels (see  \cite{chicheportiche2014fine,blanc2017quadratic} for the corresponding univariate case).

\subsection{Two- and Three-point Covariances}

The first quantity of interest is the covariance of the activities of the process.  For all $\tau\neq0$:
\begin{equation}\label{eq:defOfC}
    \mathbb{C}_{ij}(\tau) := 
    \mathop{\mathbb{E}}\bigg(\frac{{\rm d}N^i_t}{{\rm d}t}\frac{{\rm d}N^j_{t-\tau}}{{\rm d}t}\bigg) - \Bar{\lambda}_i\Bar{\lambda}_j .
\end{equation}
As for the univariate QHawkes, $\mathbb{C}$ is even, and its  extension to $\tau=0$ can be worked out following \cite{hawkes1971spectra}.
Thus, for $i=j$, the extension will be the same as in \cite{blanc2017quadratic}, with
$\mathbb{C}_{ii}^\star(\tau) := \mathbb{C}_{ii}(\tau) + \delta_{\tau,0}\Bar{\lambda}_i$. (Note indeed that $\mathop{\mathbb{E}}(({\rm d}N^i)^2)=\mathop{\mathbb{E}}({\rm d}N^i)=\Bar{\lambda}_i{\rm d}t$, if we consider that events cannot overlap). For $i\neq j$, the extension must account for co-jumps and now reads $\mathbb{C}_{ij}^\star(\tau) := \mathbb{C}_{ij}(\tau) + \delta_{\tau,0}\Bar{\mu}_{\rm{c}}$. Without co-jumps, one has $\mathbb{C}_{ij}^\star(\tau) := \mathbb{C}_{ij}(\tau)$ for $i\neq j$.

We also define a relevant three-point correlation structure $\boldsymbol{\Db}$ as the following tensor:
\begin{align}\label{eq:Ddef}
    \boldsymbol{\Db}_{ijk}(\tau_1,\tau_2) = \mathop{\mathbb{E}}\left[\left(\frac{{\rm d}N^i_t}{{\rm d}t}- \Bar{\lambda}_i \right)\frac{{\rm d}P^j_{t-\tau_1}}{{\rm d}t}\frac{{\rm d}P^k_{t-\tau_2}}{{\rm d}t}\right].
\end{align}

Since price returns are assumed to be of martingales,  $\boldsymbol{\Db}_{ijk}(\tau_1,\tau_2)$ is only nonzero when $\tau_1>0$ and $\tau_2>0$. Note that when $\tau_1=\tau_2$ and $j=k$, one has $\boldsymbol{\Db}_{ijj}(\tau,\tau) = \mathbb{C}_{ij}(\tau)$.

In the bivariate case, $\boldsymbol{\Db}$ defines again two types of $2 \times 2$ matrices: ``diagonal'' ($j=k$) and ``cross'' ($j \neq k$). We shall consistently use the following notations to distinguish them:
\begin{align}
   \mathbb{D}_{\text{d}}(\tau_1,\tau_2):=\begin{pmatrix}
    \Db_{111}(\tau_1,\tau_2) & \Db_{122}(\tau_1,\tau_2)\\
    \Db_{211}(\tau_1,\tau_2) & \Db_{222}(\tau_1,\tau_2)
    \end{pmatrix},
\end{align}
and
\begin{align}
    \mathbb{D}_{\times}(\tau_1,\tau_2)=\begin{pmatrix}
    \Db_{112}(\tau_1,\tau_2) & \Db_{121}(\tau_1,\tau_2)\\
    \Db_{212}(\tau_1,\tau_2) & \Db_{221}(\tau_1,\tau_2)
    \end{pmatrix}.
\end{align}

\subsection{Two-point Yule-Walker Equations}\label{sec:YW}

In order to deduce kernels from empirical correlations, direct relations must be determined. The method we use to find such relations is very similar to that used in Appendix 1 of \cite{blanc2017quadratic}. Assuming that prices are martingales,\footnote{This assumption is often violated at high frequencies, and some amendments will need to be introduced when calibrating the model on actual HF data -- see our Part II companion paper~\cite{ustocome}.} and without considering co-jumps, we find the following matrix equation for $\mathbb{C}$:
\begin{equation}\label{eq:C_YuleWalker_mQH}
\begin{aligned}
    &\mathbb{C}(\tau)= {\mathbb{K}_{\text{d}}}(\tau)\, \overline{\bblambda}+\int_{0^+}^{+\infty}{\mathbb{K}_{\text{d}}}(u, u)\mathbb{C}(\tau-u){\rm d}u \\
    &+2\int_{0^+}^{+\infty}\int_{u^+}^{+\infty}{\mathbb{K}_{\text{d}}}(\tau+u,\tau+r){\mathbb{D}_{\text{d}}}(u,r){\rm d}r{\rm d}u\\
    &+\int_{0^+}^{+\infty}\int_{u^+}^{+\infty}\mathbf{K}_{\cross}(\tau+u,\tau+v)(\mathbf{D}_{\cross})^\top(u,v){\rm d}v{\rm d}u\\&+\int_{0^+}^{+\infty}\int_{v^+}^{+\infty}\mathbf{K}_{\cross}(\tau+u,\tau+v)(\mathbf{D}_{\cross})^\top(u,v){\rm d}u{\rm d}v,
\end{aligned}
\end{equation}
where $\overline{\bblambda}$ is a $2 \times 2$ matrix defined as 
\begin{align}
   \overline{\bblambda}:=\begin{pmatrix}
    \Bar{\lambda}_1 & 0\\
   0&  \Bar{\lambda}_2 
    \end{pmatrix}. 
\end{align}
and where
\begin{align}
    \mathbf{D}_{\cross} := \begin{pmatrix}
    \Db_{112}\\ \Db_{212}
    \end{pmatrix}
\end{align}
This Yule-Walker equation boils down to Eq.~(8) of Ref.~\cite{blanc2017quadratic} in the univariate case.

The Yule-Walker equation for $\mathbb{C}$ accounting for co-jumps can be found in Appendix \ref{app:YWD}.

\subsection{Three-point Yule-Walker Equations}\label{sec:YW3}

The full three-point Yule-Walker equations for the tensor $\boldsymbol{\Db}$ are quite intricate. In order to give a simplified version of these equations we restrict here to the case with no co-jumps, i.e. $\mu^{\text{c}} \equiv 0$. 
We find that, for $\tau_1 > \tau_2>0$:
\begin{equation}\label{eq:D_Yule_ind_mQH}
    \begin{aligned}
&\mathbb{D}_{\text{d}}(\tau_1,\tau_2)= 2\mathbb{K}_{\text{d}}(\tau_1,\tau_2) \overline{\mathbb{C}}_{\text{d}}(\tau_2-\tau_1) \\
     &+\int_{\tau_1^+}^{+\infty}\mathbb{K}_{\text{d}}(u)\mathbb{D}_{\text{d}}(\tau_1-u,\tau_2-u)\,{\rm d}u 
    \\
    &+ 2\int_{\tau_1^+}^{+\infty}\mathbb{K}_{\text{d}}(u,\tau_1) \mathbb{D}_{\text{d}}^0(u-\tau_1, \tau_2 - \tau_1)\,{\rm d}u\\
    &+\int_{\tau_1^+}^{+\infty}\mathbf{K}_{\cross}(\tau_1,u)(\mathbf{D}^1_{\cross})^\top(u-\tau_1, \tau_2 - \tau_1){\rm d}u
    \\
    &+\int_{\tau_1^+}^{+\infty}\mathbf{K}_{\cross}(u,\tau_1)(\mathbf{D}^2_{\cross})^\top(u-\tau_1, \tau_2 - \tau_1){\rm d}u,
    \end{aligned}
\end{equation}
where we have introduced the two following diagonal matrices:
\begin{align}
   \overline{\mathbb{C}}_{\text{d}}(\tau):=\begin{pmatrix}
    \mathbb{C}_{11}(\tau) +(\Bar{\lambda}_1)^2 & 0\\
    0 &  \mathbb{C}_{22}(\tau) +(\Bar{\lambda}_2)^2
    \end{pmatrix},
\end{align}
\begin{align}
   \mathbb{D}^0_{\text{d}}(\tau_1,\tau_2):=\begin{pmatrix}
    \Db_{111}(\tau_1,\tau_2) & 0\\
    0 & \Db_{222}(\tau_1,\tau_2)
    \end{pmatrix},
\end{align}
as well as the notation $\mathbf{D}_{\cross}$ for the 2-vectors:
\begin{equation*}
    \mathbf{D}^1_{\cross} := \begin{pmatrix}
    \Db_{121}\\ 0
    \end{pmatrix},  \quad  \mathbf{D}^2_{\cross} := \begin{pmatrix}
    0 \\ \Db_{212}
    \end{pmatrix}.
\end{equation*}
For the corresponding Yule-Walker equation for $\mathbb{D}_{\cross}$ see Appendix \ref{app:YWD}.

\subsection{Asymptotic Behavior with Power Law Kernels}

An interesting special case for which the Yule-Walker equations can be asymptotically solved is when kernels are decaying as power-laws, as considered in~\cite{blanc2017quadratic}. Of special interest is the relationships between exponents governing the correlation functions and the kernel exponents when $\tau \xrightarrow{}+\infty$. 

The general analysis is quite cumbersome and is relegated to Appendix \ref{app:assym_forms}. In the non-critical case $(n<1)$, the calculations are not particularly difficult and generalise the results of \cite{blanc2017quadratic} to the multivariate case. 

The critical case $(n=1)$, on the other hand, is much more subtle, in particular for the QHawkes processes.\footnote{For the linear version, one can refer  to \cite{chicheportiche2014fine} or to \cite{bacry2016estimation} where Bacry, Jaisson and Muzy use  convolution  to link the Fourier transform of $\mathbb{C}$ with that of $\mathbb{K}$.} In order to treat this case, we have generalised the method introduced by Hawkes in \cite{hawkes1971point}, which combines the Yule-Walker equations in the Fourier domain with  Liouville's Theorem to find a direct relationship between exponents. This method is recalled in Appendix \ref{app:assym_met}, from which the value of the exponents in the critical case can be derived, see Tables in Appendix \ref{app:resultsTab}. Some of the exponent values reported in \cite{blanc2017quadratic} turn out to be incorrect; the correct values can be inferred from our new results. Note that in the critical case the decay of the correlation functions cannot be faster than $\tau^{-1}$, as in the case of linear Hawkes models without ancestors (see the work of Br\'emaud \& Massouli\'e~\cite{bremaud2001hawkes}).

\section{Power-Law Tails of the Volatility Distribution} \label{sec:vol}

Here we investigate the distribution of volatility of a  MQHawkes process. We adapt the methodology of Ref.~\cite{blanc2017quadratic} to the multivariate setting, restricting for simplicity to the two assets case. Our main goal is to establish that MQHawkes lead to fat (power-law) tails for the empirical intensity distribution, which translates into fat tails in the distribution of returns~\cite{blanc2017quadratic}. We limit our study to the ZHawkes specification with exponentially decaying kernels, that allow one to construct  a 
tractable continuous time limit.

\subsection{ZHawkes Model with Exponential Kernels}

We neglect the leverage feedback and set $\mathbb{L}=0$. We also neglect the possible presence of co-jumps, i.e. set $\mu_{\text{c}}   \equiv 0$ hereafter. Within the ZHawkes specification, we can rewrite the intensities as follows:
\begin{align*}
\begin{cases}
    \lambda_1 &= H^1_1 + H^1_2 + (Z^1_1)^2 + (Z^1_2)^2 + Y^1_2\\
    \lambda_2 &= H^2_1 + H^2_2  + (Z^2_1)^2 + (Z^2_2)^2 + Y^2_1
\end{cases}
\end{align*}
with
\begin{align*}
    H^i_j &:= \int^t_{-\infty} (\phi_{\text{d}})^i_{jj}(t-s){\rm d}N^j_s
\end{align*} 
and
\begin{align*} 
    Z^i_j &= \int^t_{-\infty} k^i_{jj}(t-s){\rm d}P^j_s\\
    Y^i_j &= \left(\int^t_{-\infty} k^i_{ji}(t-s){\rm d}P^i_s\right)\left(\int^t_{-\infty} k^i_{ij}(t-u){\rm d}P^j_u\right).
\end{align*}


To keep things tractable, we choose all kernels to be exponentials and consider that only four ``features'' are important to describe all feedback effects, namely:
\begin{align}
    \begin{cases}
     h_1(t) = {\beta_1} \int_{-\infty}^t e^{-\beta_1 (t-s)} {\rm d}N_s^1 \\
     h_2(t) = {\beta_2} \int_{-\infty}^t e^{-\beta_2 (t-s)} {\rm d}N_s^2,
    \end{cases}
\end{align}
for activity feedback, and
\begin{align}
    \begin{cases}
     z_1(t) = {\omega_1} \int_{-\infty}^t e^{-\omega_1 (t-s)} {\rm d}P_s^1 \\
     z_2(t) = {\omega_2} \int_{-\infty}^t e^{-\omega_2 (t-s)} {\rm d}P_s^2,
    \end{cases}
\end{align}
for trend feedback, with $\omega_i$'s and $\beta_i$'s positive constants. From such features, we construct the quantities $H, Z$ and $Y$ as
\begin{align*}
    H_1^i = n_{H,1}^i h_1(t)&&
    H_2^i = n_{H,2}^i h_2(t)\\
    Z^i_1 = a_{Z,1}^i z_1(t)&&
    Z^i_2 = a_{Z,2}^i z_2(t)\\
    Y^1_2 = a_{Z,\cross}^1 z_1(t) z_2(t)&&
    Y^2_1 = a_{Z,\cross}^2 z_1(t) z_{2}(t).\\
\end{align*}
Under such hypotheses, the intensities write: 
\begin{equation}\label{eq:lambda_n_a}
    \begin{aligned}
         \lambda_{1,t} =&\lambda_{1,\infty}+ n_{H,1}^1 h_1(t) + n_{H,2}^1 h_2(t)+ (a_{Z,1}^1 z_1(t))^2\\&  + (a_{Z,2}^1 z_2(t))^2 + a_{Z,\cross}^1 z_{2}(t)z_1(t)\\
        \lambda_{2,t} =&\lambda_{2,\infty}+ n_{H,1}^2 h_1(t) + n_{H,2}^2 h_2(t)+ (a_{Z,1}^2 z_1(t))^2 \\&  + (a_{Z,2}^2 z_2(t))^2 + a_{Z,\cross}^2 z_{2}(t)z_1(t).
    \end{aligned}
\end{equation}

\subsection{Endogeneity Ratios}

As mentioned earlier, the Hawkes endogeneity ratio $n_H$ is obtained as the spectral radius of the $2 \times 2$ matrix of Hawkes coefficients, namely
\begin{align}
    \mathbb{N}_{H}:=\begin{pmatrix}
    n_{H,1}^1 & n_{H,2}^1\\
    n_{H,1}^2 & n_{H,2}^2
    \end{pmatrix}.
\end{align}

The Zumbach endogeneity coefficient $n_Z$ is the spectral radius of the trend feedback kernel. 
In the monovariate case, one finds $n_Z=\int_{0}^{+\infty}k^2(s){\rm d}s$. From the features defined above, we now have $k^i_{jj}(t)=a^i_{Z,j}\omega_je^{-\omega_jt}$. Noting that 
\[ 
\omega^2 \int_0^\infty e^{-2 \omega s}\, {\rm d}s = \frac{\omega}{2}, 
\]
we simply need, in the present case, to find the top eigenvalue of the following $2 \times 2$ matrix: 
\begin{align}
    \mathbb{N}_{Z}:=\frac12 \begin{pmatrix}
    (a_{Z,1}^1)^2 \omega_1 & (a_{Z,2}^1)^2 \omega_2\\
    (a_{Z,1}^2)^2 \omega_1 & (a_{Z,2}^2)^2 \omega_2
    \end{pmatrix}.
\end{align}
In the simplest case when all coefficients and decay rates are equal, one finds that $n_H$ is equal to any of the coefficients $n_{H,j}^i$ (which are all equal), and $n_Z = a_Z^2 \omega$, where $a_Z:=a_{Z,j}^i$ (again all equal).

Note that in the general case, one needs to diagonalize the matrix $\mathbb{N}_H + \mathbb{N}_Z$, which leads to a total endogeneity ratio that is in general different from $n_H + n_Z$.

\subsection{Fokker-Planck Equation }

As in \cite{blanc2017quadratic}, we consider the process on a time scale $T>0$, that shall eventually tend to $+\infty$, and simultaneously take all decay rates $\beta, \omega \to 0$, but in such a way that $\omega T$ and $\beta T$ remain finite. For the Zumbach feedback not to disappear in this limit one  needs to simultaneously scale up both $(a_{Z,j}^i)_{i,j \in \{1,2\}}$ as $\omega^{-1/2}$ and $(a_{Z,\cross}^i)_{i \in \{1,2\}}$ as $\omega^{-1}$. In this scaling regime, one can establish the following Fokker-Planck equation for the time dependent probability density $p_t$ of $(h_1,h_2,z_1,z_2)$ (see Appendix \ref{app:infiniV2}): 

\begin{equation}\label{eq:FP_gen}
    \begin{aligned}
        \partial_t p_t =&\,
     {\beta}_{1}\partial_{h_1}\big((h_1-\lambda_{1,t})p_t\big)
 +{\beta}_{2}\partial_{h_2}\big((h_2-\lambda_{2,t})p_t\big)\\  
 &+{\omega}_{1}\partial_{z_{1}}\big(z_{1}p_t \big)
 +{\omega}_{2}\partial_{z_{2}}\big(z_{2}p_t\big) \\
 &+\frac{\omega_1^2}{2}\partial^2_{z_1z_1}\big(\lambda_{1,t} p_t\big)
+\frac{\omega_2^2}{2}\partial^2_{z_2z_2}\big(\lambda_{2,t} p_t\big),
    \end{aligned}
\end{equation}
where $p_t$ is a shorthand for $p_t(h_1,h_2,z_1,z_2)$, $\lambda_{1,t}$ and $\lambda_{2,t}$ are given by Eq.~\eqref{eq:lambda_n_a}, and where we have disregarded co-jumps and direct correlations between the returns of asset 1 and asset 2, meaning that $\mathbb{E}({\rm d}P^1 {\rm d}P^2)=0$. The inclusion of such correlations can be considered and adds further cross-derivative terms $\partial^2_{z_1z_2}$ in the Fokker-Planck equation.

Solving for the stationary distribution $p_\infty$ of the Fokker Planck equation allows one to determine the tail behaviour of the distribution of intensities (which translate into volatilities since $\mathbb{E}[({\rm d}P_t)^2] = \lambda_t {\rm d}t$). In the monovariate case, Ref.~\cite{blanc2017quadratic} established that $p_\infty$ decays as a power-law, with an exponent $\alpha$ that depends  on both $n_H$ and  $n_Z$. The general expression for $\alpha$ is however not available in closed form, although asymptotic results in various regimes could be worked out, in particular when $n_H \to 0$. 
The most important conclusion is that $\alpha \to \infty$ when $n_Z \to 0$, i.e. power-law tails disappear in the absence of a quadratic Zumbach coupling. Interestingly, the coupling between the Hawkes feedback with $n_H \sim 1$ and even a small Zumbach effect ($n_Z \ll 1$) was shown to generate an exponent compatible with empirical data.

\subsection{ZHawkes without Hawkes ($n_H=0$)}

 To make further analytical progress, we now consider the case in which the Hawkes coupling is absent ($n_H=0$), that is when $h$ and $z$ decouple, leading to a tractable two-dimensional Fokker-Planck for $z_1$ and $z_2$.
 In the stationary regime, the Fokker Planck equation~\eqref{eq:FP_gen} becomes:
\begin{equation}\label{eq:FP_Hawkes_without_Hawkes}
    \begin{aligned}
    &0 = {\omega}_{1}\partial_{z_{1}}\big(z_{1}p_\infty(z_1,z_2) \big)
 +{\omega}_{2}\partial_{z_{2}}\big(z_{2}p_\infty(z_1,z_2)\big) \\
 &+\frac{\omega_1^2}{2}\partial^2_{z_1z_1}\big(\lambda_1 p_\infty(z_1,z_2)\big)
+\frac{\omega_2^2}{2}\partial^2_{z_2z_2}\big(\lambda_2 p_\infty(z_1,z_2)\big).
    \end{aligned}
\end{equation}
This equation describes the stationary measure of the stochastic path of the bivariate process $(Z_1,Z_2)_t$ defined as: 
\begin{equation}
\begin{aligned}
    & {\rm d}Z_1 =-{\omega}_1 Z_1 {\rm d}t+ {\omega}_1\sqrt{\lambda_{1,t}} {\rm d}W^1_t\\
  &  {\rm d}Z_2 =-{\omega}_2 Z_2 {\rm d}t+ {\omega}_2 \sqrt{\lambda_{2,t}} {\rm d}W^2_t, \nonumber
\end{aligned}
\end{equation}
with
\begin{equation}\label{eq:ito_path}
\begin{aligned}
  &   \lambda_{1,t} =\lambda_{1,\infty}+a_{Z,\cross}^1 Z_{2}Z_1 + (a_{Z,1}^1 Z_1)^2 + (a_{Z,2}^1 Z_2)^2  \\
 &   \lambda_{2,t} =\lambda_{2,\infty}+ a_{Z,\cross}^2 Z_{2}Z_1+ (a_{Z,2}^2 Z_2)^2 + (a_{Z,1}^2 Z_1)^2. 
\end{aligned}
\end{equation}
These equations allow one to simulate paths $(Z_1,Z_2)_t$ numerically, from which an empirical determination of $p_\infty(z_1,z_2)$ can be confronted to our analytical solution of Eq.~\eqref{eq:FP_Hawkes_without_Hawkes}. Note that the coefficients $(a_{Z,j}^i)_{i,j \in \{1,2\}}$ and $(a_{Z,\cross}^i)_{i \in \{1,2\}}$ must satisfy the following inequalities for $\lambda_{1,t}$ and $\lambda_{2,t}$ to remain positive at all times:
\[ 
4 (a_{Z,1}^i a_{Z,2}^i)^2 \geq (a_{Z,\cross}^i)^2, \qquad i=1,2.
\]
How can one determine the tail exponent for such a two dimensional process? We first introduce polar coordinates $(r,\theta)$, such that $z_1=r\cos\theta $ and $z_2=r\sin\theta $. We then surmise that when $r^2 \gg \max(\lambda_{1,\infty}, \lambda_{2,\infty})$ the stationary distribution $p_\infty(r,\theta)$ factorizes into an angular component $F(\theta)$ and a power-law contribution, to wit:  $p_\infty(r,\theta) \underset{r \to \infty}{\approx} {F(\theta)}{r^{-\alpha}}$,  where $\alpha$ is the tail exponent. Note that $\alpha$ should be strictly larger than 2 for $p_\infty$ to be normalisable, and strictly larger than 3 for the mean intensity to be non divergent. Injecting our factorized guess into Eq.~\eqref{eq:FP_Hawkes_without_Hawkes}  and taking the limit $r^2\gg\lambda^{\infty}$, we find a second order ODE on the function $F$, where $\alpha$ appears as a parameter (see Appendix \ref{app:ODE_F}, Eq.~\eqref{eqApp:ODEonF}).

The value of $\alpha$ is selected by the analogue of energy quantification in quantum mechanics: only for some special values of $\alpha$ can one find a solution $F$ of the above ODE that satisfies the correct boundary conditions compatible with symmetries of the problem. Clearly, $F$ must be everywhere non-negative and, because of the symmetry $z_1 \to -z_1$ and $z_2 \to -z_2$, one must have $F(\theta+ \pi)=F(\theta)$ (see below for explicit examples). In principle, there can be more than one value of $\alpha$ that allows one to find an acceptable solution $F$. This is the analogue of the energy spectrum in quantum mechanics. The value of $\alpha$ that governs the tail behaviour of $p_\infty(r,\theta)$ is then the \textit{smallest} of all such acceptable values.   
Once the asymptotic tail behaviour of $p_\infty(r,\theta)$ is known, it is easy to derive the tail behaviour of the marginals $p_\infty(z_1)$ and $p_\infty(z_2)$, which both behave as $z^{-1-\mu}$ with $\mu = \alpha - 2$. The volatility distribution then has the same tail behaviour. 

In order to illustrate this general procedure on a simple example, we focus in the following on the case where  $a_{Z,i}^j = \sqrt{2 n_Z/\omega_i}$, $\forall i,j = 1,2$, such that the two eigenvalues of $\mathbb{N}_Z$ are equal to $n_Z$ (the Zumbach endogeneity ratio) and $0$. We also set $a_{Z,\cross}^i = 2 \gamma n_Z/\sqrt{\omega_1 \omega_2}$, where $\gamma$ is an arbitrary coefficient $\in (-2,2)$ (such that $\lambda_1$ and $\lambda_2$ are always positive). When $\omega_1=\omega_2$,  Eq.~\eqref{eqApp:ODEonF} considerably simplifies and reads:

\begin{equation}\label{eq:Ftheta}
    \begin{aligned} 
    &\left[(\alpha-2)\left(\alpha-\alpha_0\right)+(\alpha-2)^2\gamma  \cos(\theta) \sin(\theta)  \right]F(\theta)  \\ & +\left[(1+\gamma\cos(\theta) \sin(\theta) )F(\theta)\right]''=0,
\end{aligned}
\end{equation}
where we have defined 
\begin{equation*}
    \alpha_0:= 2 + \frac{1}{n_Z}.
\end{equation*}
Note that for a given value of $\alpha$ this equation is invariant under the simultaneous change $\gamma \to - \gamma$, $\theta \to - \theta$. Hence the value of $\alpha$ can only depend on $|\gamma|$.

\subsubsection{The isotropic case $\gamma = 0$}\label{sec:isotropiccase}

When cross terms  $a_{Z,\cross}^i$ are absent and $\omega_1=\omega_2$, the problem becomes isotropic in the sense that the dynamics of $r^2 = z_1^2 + z_2^2$ decouple from that of $\theta$. The problem then boils down to the univariate ZHawkes model without Hawkes coupling, for which the value of $\alpha$ is known, and given by $\alpha_0$.

Furthermore, the evolution of $\theta$ is that of a free Brownian motion on the unit circle, leading to a uniform distribution $F(\theta)=F_0$, which is indeed a solution of Eq.~\eqref{eq:Ftheta} in this case. Note that other periodic solutions exist whenever 
\[ 
(\alpha-2)(\alpha-\alpha_0) = 4 \ell^2, \qquad \ell=0,1,2, \ldots
\]
but lead to larger values of $\alpha$ when $\ell > 0$.

\subsubsection{The case $\gamma \neq 0$}
\label{sec:anisotropic}
In order to make progress, we posit that $\alpha$ and $F$ can be expanded as  power series of $\gamma$, namely
\begin{equation*}
    \begin{aligned}
    &    \alpha=\alpha_0 + \alpha_1\gamma +\alpha_2\gamma^2 +\ldots\\
     &   F(\theta) = F_0 + F_1(\theta)\gamma + F_2(\theta)\gamma^2+ \ldots
    \end{aligned}
\end{equation*}
where $\alpha_0=2+{1}/{n_Z}$ is the solution for $\gamma=0$ and $F_0$ is the constant solution found above. The coefficient $\alpha_1$ must be zero for symmetry reasons.

Inserting this expansion in Eq.~\eqref{eq:Ftheta} and imposing that all $F_n(\cdot)$ remain $\pi$-periodic, the identification of terms of order $\gamma^n$  finally leads to
\begin{equation}\label{eq:small_gamma}
    \alpha = \alpha_0 +\frac{\gamma^2}{32}\left(\frac{4}{n_Z}-\frac{1}{n_Z^3}\right) + O(\gamma^4).
\end{equation}

\begin{figure}
    \centering
    \includegraphics[width=\linewidth]{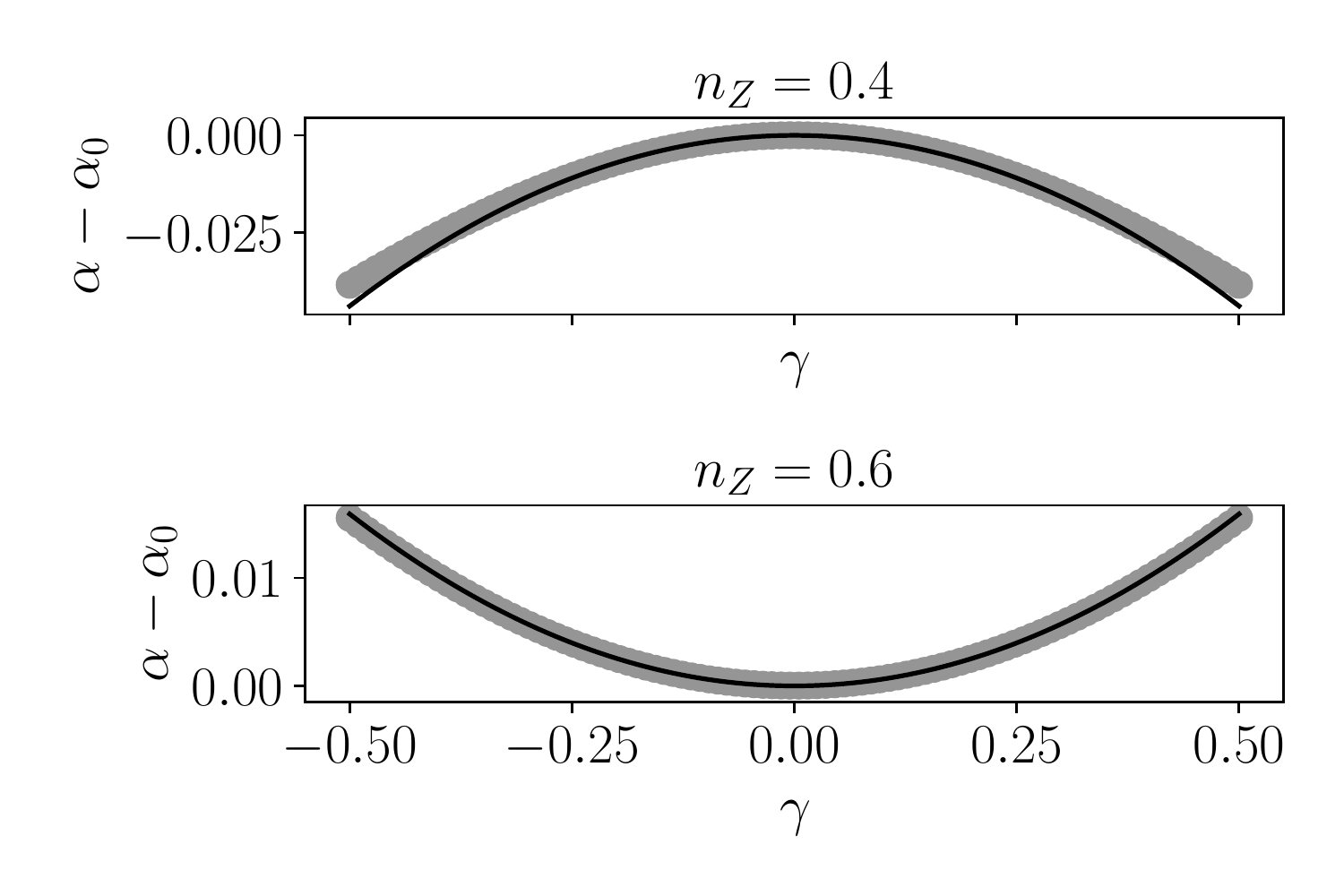}
    \caption{Plot of $\alpha-\alpha_0$ as function of $\gamma$ for $n_Z=0.4$ (above) and $n_Z=0.6$ (below). The $\alpha$ are found numerically by looking for the smallest $\alpha$ for a  $\pi$-periodic and positive solution $F$. The solid black line represents the theoretical prediction,  Eq.~\eqref{eq:small_gamma}, that is $\alpha \approx \alpha_0 -0.18\gamma^2$ for $n_Z=0.4$ and $\alpha \approx \alpha_0 +0.06\gamma^2$  for $n_Z=0.6$.}
    \label{fig:alpha_fct_gamma}
\end{figure}
Figure~\ref{fig:alpha_fct_gamma} displays the numerical values of $\alpha$ as a function of $\gamma$ for $n_Z=0.4$ and $n_Z=0.6$ in with the corresponding theoretical parabolas, see Eq.~\eqref{eq:small_gamma}.  
Note that the $\gamma^2$ correction changes sign when $n_Z=1/2$. 

In this case ($n_Z=1/2$), finding an exact solution of the associated Schrodinger solution is possible~\cite{xie2011new}, and  leads to 
$\alpha=4$ for all values of $\gamma$.\footnote{The other exact solutions found in \cite{xie2011new} unfortunately correspond to sub-dominant, larger values of $\alpha$.} The corresponding solution for $F$ is also known in that case and is a constant independent of $\theta$, as can be directly checked from Eq. ~\eqref{eq:Ftheta}. An expansion around the special point  $n_Z=1/2$ can in fact be performed and leads to first order to 
\[
\alpha=4 + \frac{16}{4+\gamma^2} \left(\frac{1}{2}-n_Z\right)+o\left(\frac{1}{2}-n_Z\right).\] 

Finally, note that the condition $\alpha > 3$ (ensuring that the mean activity is finite) reads, to first order in $\gamma$:
\begin{equation} \label{eq:n-star}
    n_Z < n^\star \approx 1 + \frac{\gamma^2}{8} + O(\gamma^4).
\end{equation}
The isotropic case $\gamma = 0$ boils down to the univariate ZHawkes model, for which $n^\star=1$, see \cite{aubrun2022hawkes}.

\begin{figure}[b!]
    \centering
    \includegraphics[width=\linewidth]{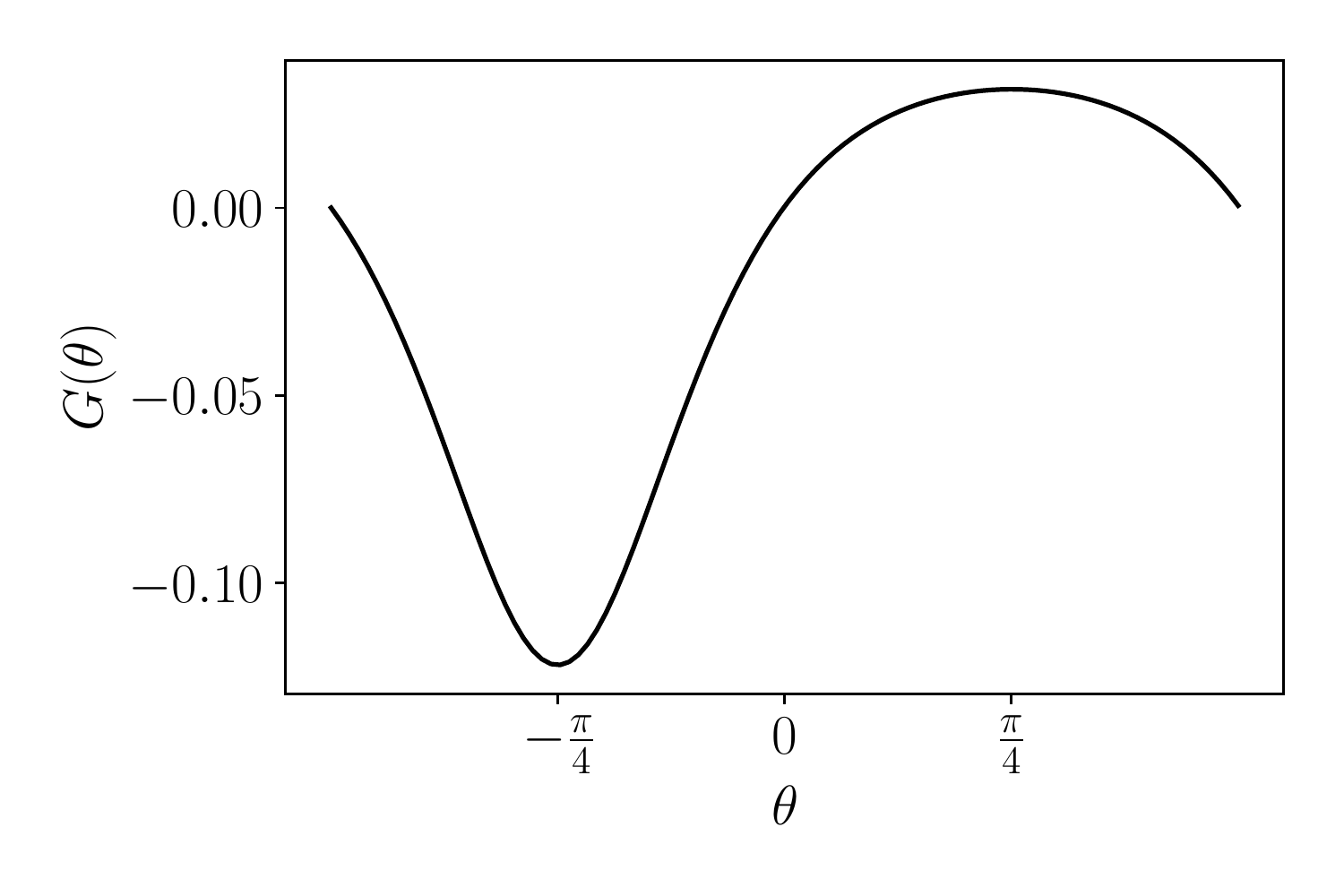}
    \caption{Plot of $G(\theta)$ for $\gamma=1$, $C=0.6$, see Eq.~\eqref{eq:Gtheta}.}
    \label{fig:G_sol}
\end{figure}

\subsubsection{The case $n_Z\xrightarrow{}+\infty$}\label{sec:limnz}

Equation~\eqref{eq:Ftheta} can be easily analysed when $n_Z\xrightarrow{}+\infty$, in which case $\alpha$ tends to the smallest possible value, $2$, corresponding to a maximally ``fat'' distribution. The only periodic solution to Eq.~\eqref{eq:Ftheta} in that limit is 
\begin{equation}
    F_\infty(\theta)=\frac{C}{2+\gamma\sin(2\theta)},
\end{equation}
where $C$ is a constant. 
When $n_Z$ is very large but not infinite, we  posit that the solution  writes
\begin{equation}\label{eq:FperturbativenZ}
    F(\theta)=F_\infty(\theta) + \frac{1}{n_Z^2} G(\theta) + O(\frac{1}{n_Z^4}),
\end{equation}
together with $\alpha = 2 + \zeta/n_Z$, with $G(\cdot)$ and $\zeta$ to be determined. Inserting Eq.~\eqref{eq:FperturbativenZ} in the ODE on $F$ Eq.~\eqref{eq:Ftheta}, one obtains the ODE for $G$:
\begin{equation}\label{eq:Gtheta}
    \begin{aligned} 
  \left[(2+\gamma\sin(2\theta))G(\theta)\right]''= - C \frac{2\zeta (\zeta-1)+ \gamma \zeta^2 \sin(2\theta)}{2+\gamma\sin(2\theta)}.
\end{aligned}
\end{equation}

Imposing $G(\cdot)$ to be $\pi$-periodic, that is imposing that the right hand side of Eq.~\eqref{eq:Gtheta}  integrates to zero between $\theta=0$ and $\theta=\pi$, sets the value of $\zeta$: 
\begin{equation}
\zeta = \left(1-\frac{\gamma^2}{4}\right)^{-\frac12},
\end{equation}
recovering $\zeta=1$ when $\gamma=0$ and the small $\gamma$ expansion result above, see Eq.~\eqref{eq:small_gamma}. The function $G(\cdot)$ is plotted in Fig.~\ref{fig:G_sol} for $\gamma=1$.

\begin{figure}
    \centering
    \includegraphics[width=\linewidth]{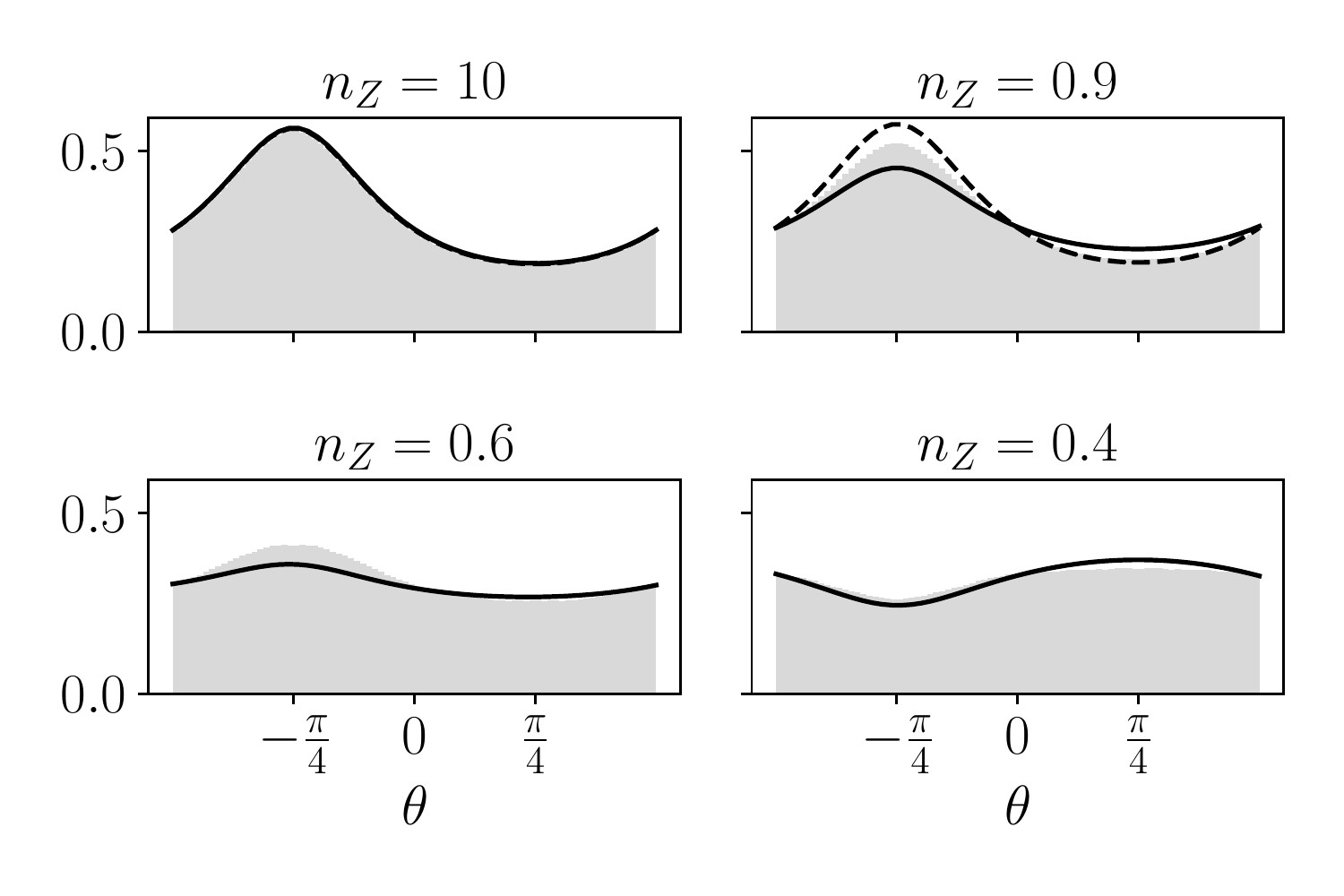}
    \caption{Comparison between the numerical solution $F$ of Eq.~\eqref{eq:Ftheta} (black lines) and the results of simulating Eq.~\eqref{eq:ito_path} (purple histogram)  for several values of $n_Z$, and $\gamma=1$. The solid line shows the numerical solution $F$ found by  looking for a solution couple $(F, \alpha)$ of a positive $\pi$-periodic solution $F$ of Eq.~\eqref{eq:Ftheta} and a value $\alpha$. The the dashed line for $n_Z=10$ and $n_Z=0.9$ shows the solution $F_\infty$ for $\alpha=2$ ($n_Z\xrightarrow{}+\infty$). The upper right figure, obtained for $n_Z=10$, matches very well the solution obtained when $n_Z\xrightarrow{}+\infty$ since corrections are $O(n_Z^{-2})$. Note that when $n_Z=1/2$, the exact solution $F(\theta)$ is independent of $\theta$ for all values of $\gamma$.}
    \label{fig:comparisonSolutionsHistTheta}
\end{figure}

When comparing the solution above with the histogram of simulated $\theta$ for $n_Z=10$ and $\gamma=1$, we find an excellent match with $F_\infty$, without need of any correction, see  Fig.~\ref{fig:comparisonSolutionsHistTheta}. This is expected since the correction term $G(\theta)/n_Z^2$ is of the order of $1 \%$ in that case.
When $n_Z$ decreases, corrections become more pronounced. The numerically computed $F$ is in good agreement with the angular distribution obtained from a direct numerical simulation of the two dimensional stochastic process (Eq.~\eqref{eq:ito_path}).
 
\subsubsection{The case $n_Z\xrightarrow{}0$}

When $n_Z$ tends to zero, we expect that the exponent $\alpha$ of the power-law tail diverges. Looking again for  $\alpha$ of the form $\alpha = 2 + \zeta/n_Z$, we find that Eq.~\eqref{eq:Ftheta} reads:
\begin{equation}\label{eq:Ftheta3}
\begin{aligned}
    &\big(2\zeta(\zeta-1)+\zeta^2\gamma\sin(2\theta) \big)F(\theta)  \\ & +n_Z^2 \left[(2+\gamma\sin(2\theta))F(\theta)\right]''=0.
\end{aligned}
\end{equation}

\begin{figure}
    \centering
    \includegraphics[width=\linewidth]{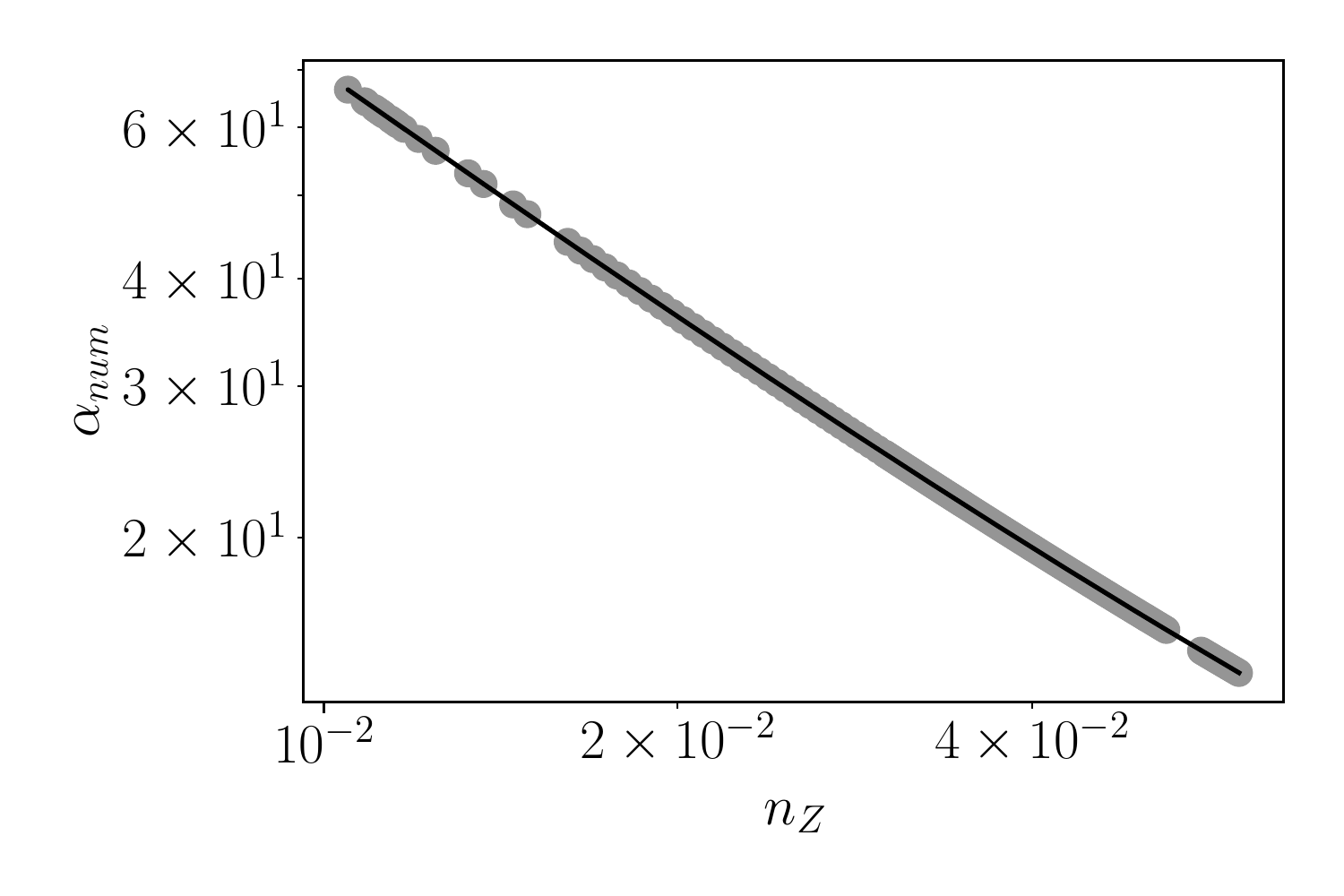}
    \caption{Tail exponent $\alpha$ as a function of $n_Z$ for $n_Z\leq0.06$. Circles:  numerical determination of $\alpha$ such that $F$ is a $\pi$-periodic solution of equation~ \eqref{eq:Ftheta}. Solid black line: theoretical value  $\alpha=2+{\zeta}/{n_Z}$ with $\zeta$ given by Eq.~\eqref{eq:zeta_nz_tend_0} }
    \label{fig:alpha_nz_tend_0}
\end{figure}

When $n_Z\xrightarrow{}0$, this equation looks self-contradictory because the remaining term can only be zero if $F(\theta)=0$. However, the second derivative term is a singular perturbation, so it must be treated with care. The idea is to look for a solution $F$ which is zero nearly everywhere, except very close to some special values of $\theta$ where the second derivative diverges. 
It turns out that all the action takes place close to $\theta=\pi/4$ when $\gamma > 0$ and $\theta=-\pi/4$ when $\gamma < 0$. Choosing $\gamma > 0$, $\theta=\pi/4 + u \sqrt{n_Z}$, and $n_Z\xrightarrow{}0$, Eq.~ \eqref{eq:Ftheta3} becomes the Schrodinger equation of a harmonic oscillator, up to terms $O(n_Z^2)$:
\begin{equation*} 
\left[ - ( 2 + {\gamma}) \frac{{\rm d}^2 \Psi}{{\rm d}u^2} +  2\zeta^2 \gamma u^2 \Psi \right] =   \frac{2\zeta(\zeta-1)+{\zeta^2 \gamma}}{n_Z}\Psi,
\end{equation*} 
with 
\[
\Psi(u):=F\left(\frac{\pi}{4} + {u}{\sqrt{n_Z}}\right).
\]
\begin{figure}[b!]
    \centering
    \includegraphics[width=\linewidth]{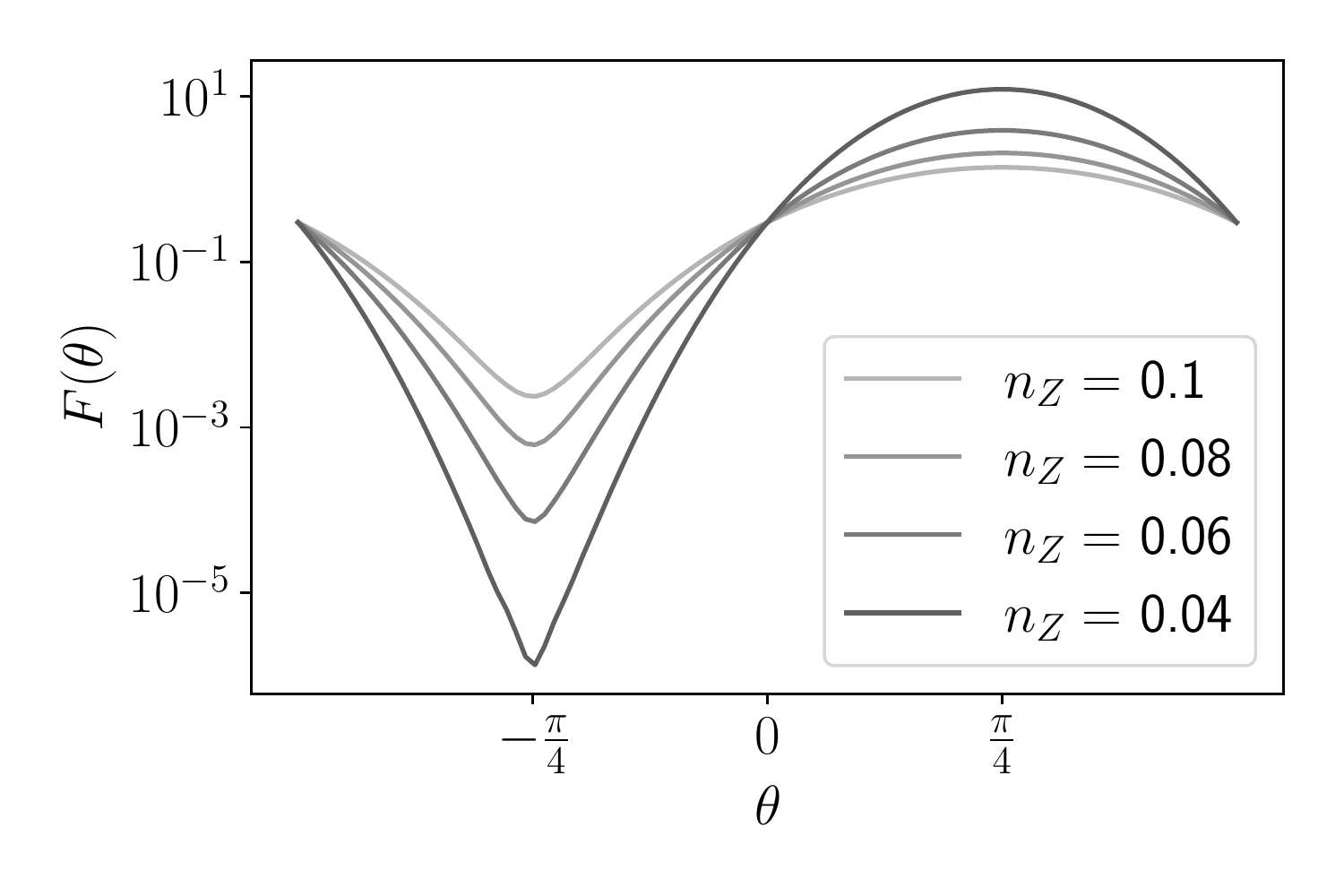}
    \caption{Solution $F$ of Eq.~\eqref{eq:Ftheta} when $n_Z \leq 0.1$. Note the logarithm scale on the y-axis, such that the Gaussian solution for $\Psi(u)$, Eq.~\eqref{eq:Psi}, is a parabola around $\theta=\pi/4$.}
    \label{fig:sol_nz_goes_to_0}
\end{figure}
The smallest $\alpha$ solution (or ``ground state'') is
\begin{equation} \label{eq:Psi}
\Psi(u)= C' e^{-\varkappa u^2} + O(n_Z), \qquad \varkappa = \zeta \sqrt{\frac{\gamma}{4 + 2\gamma}}, 
\end{equation}
where $C'$ is another constant,
together with 
\begin{equation}\label{eq:zeta_nz_tend_0}
    \zeta = \frac{2}{2 + \gamma} + \sqrt{\frac{2\gamma}{2 + \gamma}} n_Z + O(n_Z^2).
\end{equation}
This solution is only accurate in a region of width $\sim\sqrt{n_Z}$ around $\pi/4$, beyond which it quickly goes to zero. This is in perfect agreement with the numerical solution of Eq.~\eqref{eq:Ftheta} for small $n_Z$, shown in Fig.~\ref{fig:sol_nz_goes_to_0}. The parabolic shape in a semi-log plot around $\theta=\pi/4$ agrees with the predictions of Eq.~\eqref{eq:Psi}. Moreover, the value of $\alpha=2+{\zeta}/{n_Z}$, with $\zeta$ given in Eq.~\eqref{eq:zeta_nz_tend_0}, also perfectly matches the numerical values reported in Fig.~\ref{fig:alpha_nz_tend_0}. Note that for $\gamma < 0$, the same results hold with $|\gamma|$ replacing $\gamma$ in the above equations.

\subsection{The General Case}

In the previous section, we have shown how to compute the power-law tail exponent in the case where only the quadratic ``ZHawkes'' kernel is present. We also restricted to simple cases where the frequencies $\omega$ and coupling constants $a_Z$ are symmetric. Although analytically more challenging, the method outlined above can be implemented more generally, and amounts to solving a problem akin to the quantification condition for the Schrödinger equation. Similarly to the monovariate case, the volatility distribution develops a power-law tail for all values of the Hawkes feedback  $n_H$, as long as some amount of Zumbach feedback $n_Z$ is present. When $n_H \neq 0$, the equation setting the tail in the bivariate case is a three dimensional partial derivative equation generalizing the equation written in the appendix of Ref.~\cite{blanc2017quadratic}. We expect that even a tiny amount of Zumbach feedback coupled to the standard Hawkes effect brings the exponent $\alpha$ into the empirical range, as found in the monovariate case~\cite{blanc2017quadratic}.

\section{Conclusion}
\label{sec:conclusion}

Let us summarise what we have achieved in this study. Building on the work of Blanc \textit{et al.}~\cite{blanc2017quadratic}, we extended the Quadratic Hawkes model to a multivariate framework (MQHawkes). We emphasized that in the multivariate case, both idiosyncratic and common jumps must be considered in the general case, leading to a quite complex framework that we only detailed in the bivariate case. 

We defined the endogeneity ratio of MQHawkes, as well as the associated conditions for the process to be stationary. Within the ZHawkes approximation -- where quadratic kernels write as a sum of a time diagonal component, reproducing a linear Hawkes feedback, and a rank one component --  we gave a deeper understanding of the roles of the different feedback terms in the stationarity condition. We found that, in particular, the spectral radius of the Hawkes component needs to be strictly inferior to 1, as for the 1D case. The rank one component contains both exciting and inhibiting realisations and is not involved in the condition, although such a contribution can lead to a divergence of the average intensity of the process, see \cite{aubrun2022hawkes}. 

We further defined the covariance structures for MQHawkes processed and established the associated Yule-Walker equations. The latter allow one to fully determine the quadratic kernels from data, and thereby pave the way for their empirical calibration.

Finally, we studied the volatility distribution of a 2D MQHawkes process. Restricting our study to ZHawkes without Hawkes $(n_H=0)$, with exponential kernels and in symmetric cases, we were able to characterize exactly the tail of the joint probability density function of the ZHawkes intensity terms. We found that it displays a power law behavior, in line with the observed fat tails of financial returns. Note that, interestingly, the coupling between assets imposes that the exponent $\alpha$ is the same for all assets -- a mechanism that may explain the apparent universality of the power-law tail observed in financial markets. 

In a forthcoming companion paper \cite{ustocome}, we shall calibrate the model on empirical data. On the theoretical side, while expected to be quite heavy, it would be interesting to further develop the analysis in the presence of co-jumps and correlations. Note that for the sake of clarity most of our explicit expressions are given in the 2D case. Expanding them further in the $N$-dimensional case will be necessary in order to calibrate the model on a large number of assets, say a pool of stocks \cite{ustocome}.

Also, in the present analysis of the volatility distribution, we restricted to symmetric coefficients and exponential kernels. Confirming that our conclusions  are qualitatively robust against changes in symmetry and kernel functionnals, and studying its quantitative implications would also be of interest. Furthermore, we have focused on the specific case of the quadratic Hawkes model because of its ability to reproduce multiple financial stylised facts, but it would also be interesting to consider the multivariate generalisation of other non-linear Hawkes models, following the interesting insights and methods presented in Ref. \cite{kanazawa2021ubiquitous}.

\section*{Acknowledgements} \label{sec:acknowledgements}
We would like to thank Riccardo Marcaccioli for sharing his work and ideas on detection of jumps in financial data, Natascha Hey and Swann Chelly for fruitful discussions and Jérôme Garnier-Brun, Rudy Morel and Samy Lakhal for their insights on numerical simulations. 
  
  This research was conducted within the
Econophysics \& Complex Systems Research Chair, under the aegis of the Fondation du Risque, the Fondation de l’Ecole polytechnique, the Ecole polytechnique and Capital Fund Management.

\clearpage

\bibliographystyle{vancouver}
\bibliography{biblio}

\clearpage

\newpage
\appendix*

\small 
\onecolumngrid
\subsection{Notations}\label{app:notations}
The notations and formulae described on this page will be used throughout the paper.
\begin{align*}
&\mathbb{A} & &\text{Matrix}\\
&\mathbb{A}^\top & &\text{Matrix transpose}\\
&\mathbb{A}(\tau,\tau) & &\text{Matrix time diagonal}\\
&\mathbb{A}_{ii} & &\text{Matrix diagonal}\\
&\mathbf{X} & &\text{Vector}\\
&(N^i_t)_{t\geq0} & &\text{Quadratic Hawkes process describing the time of the price changes of asset $i$}\\
&(N^c_t)_{t\geq0} & &\text{Quadratic Hawkes process describing the times of co-jumps for 2 price processes in the correlated case}\\
&(P^i_t)_{t\geq0} & &\text{Price process of the asset $i$}\\
&\psi & &\text{Tick size}\\
&\lambda_{i,t} & &\text{Intensity of the QHawkes process $N^i$ at time $t$}\\
&\lambda_{i,\infty} & &\text{Baseline intensity of the QHawkes process $N^i$}\\
&\mu_{i,t} & &\text{Intensity of the QHawkes process $N^i$ at time $t$ in the correlated case}\\
&\mu_{i,\infty} & &\text{Baseline intensity of the QHawkes process $N^i$ in the correlated case}\\
&n/ n_H/ n_Q/ n_Z & &\text{Endogeneity ratio of the process/Hawkes contribution/regular contribution/Zumbach  contribution}\\
&\epsilon_t & &\text{Sign of the price change at time $t$, $\epsilon_t \in \{-1,1\}$}\\
&L_j^i & &\text{Leverage kernel of intensity $i$ reflecting the feedback of the price changes of asset $j$}\\
&K_{jk}^i & &\text{Quadratic kernel of intensity $i$ reflecting the feedback of the price changes of assets $j$ and $k$}\\
&Q_{jk}^i & &\text{Same as $K_{jk}^i$ for the correlated case}\\
&\rho & &\text{Correlation of the signs of the price changes ${\rm d}P^1$ and ${\rm d}P^2$ in the correlated case}\\
&\mathbb{K}_{\text{d}} & &\text{Kernel matrix such that }{\mathbb{K}_{\text{d}}}_{ij}=K^i_{jj}\\
&\mathbf{K_{\times}} & &\text{Kernel vector representing cross feedback, composed by $K^i_{jk}$ with $j<k$}\\
&\boldsymbol{\Bar{\lambda}} & &\text{Mean Intensity}\\
&\delta_{\cdot,\cdot} & &\text{Kronecker delta: }\delta_{x,y}=1,\text{ when $x=y$ and $\delta_{x,y}=0$ otherwise}\\
&\delta(\cdot) & &\text{Dirac mass}\\
&\mathbb{C}_{ij} & &\text{Covariance } \mathbb{C}_{ij}(\tau):=\mathop{\mathbb{E}}\bigg(\frac{{\rm d}N^i_t}{{\rm d}t}\frac{{\rm d}N^j_{t-\tau}}{{\rm d}t}\bigg) - \Bar{\lambda}_i\Bar{\lambda}_j\\
&\boldsymbol{\Db}_{ijk} & &\text{3 points correlation structure }\boldsymbol{\Db}_{ijk}(\tau_1,\tau_2):=\mathop{\mathbb{E}}\bigg(\frac{{\rm d}N^i_t}{{\rm d}t}\frac{{\rm d}P^j_{t-\tau_1}}{{\rm d}t}\frac{{\rm d}P^k_{t-\tau_2}}{{\rm d}t}\bigg)\\
&\mathbb{D}_{\times} & &\text{Square (2x2) correlation matrix such that }{\mathbb{D}_{\times}}_{ij}=\boldsymbol{\Db}_{ijj}\\
&\mathbb{D}_{\text{d}}& &\text{Square (2x2) correlation matrix such that }{\mathbb{D}_{\times}}_{ij}=\boldsymbol{\Db}_{ijk} \text{ with }j\neq k\\
&(\phi_{\text{d}})^i_{jj} & &\text{Time diagonal contribution of }{(\mathbb{K}_{\text{d}})}^i_{jj}\\
&k^i_{jk} & &\text{Rank one contribution of }K^i_{jk}\\
&H^i_j & &\text{Hawkes component of intensity $i$ from feedback $j$; } H^i_j=n_{H,j}^ih_j(t) \\
&n_{H,j}^i & &\text{Endogeneity ratio of Hawkes component }H^i_j\\
&Z^i_j & &\text{Quadratic component of intensity $i$ from feedback $j$}\\
&a_{Z,j}^i & &\text{Amplitude of the quadratic component }Z^i_j \text{; }a_{Z,i}^j = \sqrt{2 n_Z/\omega_i}\\
&Y_{j}^i & & \text{Cross component of intensity $i$; }Y_{j}^i=a_{Z,\cross}^iz_1z_2\\
&a_{Z,\cross}^i & &\text{Amplitude of the cross component of intensity $i$} \\
&\gamma & &\text{Anisotropy coefficient $\gamma \in (-2,2)$, such that }a_{Z,\cross}^i = 2 \gamma n_Z/\sqrt{\omega_1 \omega_2}\\
&p_\infty(r,\theta) \underset{r \to \infty}{\approx} {F(\theta)}{r^{-\alpha}}& &\text{Joint pdf of $(z_1,z_2)$ in polar coordinates in the case ZHawkes without Hawkes with exponential kernels.}
\end{align*}

\onecolumngrid


\subsection{Conditions for Positive QHawkes Intensities}\label{app:positiveLambda}


Inspired by the method in \cite{blanc2014fine}, we detail here sufficient conditions for the intensity of the process $i$ to be positive, when considering $N$ assets. We first consider the simple case $\mathbb{L}\equiv0$.   First, considering that the kernels are negligible up to a value $q$, and using a discrete approximation of the integration, we rewrite the intensity associated with asset $i$: 

\begin{equation*}
\begin{aligned}
    \lambda_{i,t} =& \lambda_{i,\infty}+\sum_{j \leq k}^N  \iint^t_{-\infty} K_{jk}^i(t-s,t-u)\, {{\rm d}P^j_s}\, {\rm d}P^k_u \simeq \lambda_{i,\infty}+\sum_{j \leq k}^N  \sum_{u=1}^q \sum_{s=1}^q {\rm d}P^j_{t-u}{K}^i_{jk}(s,u){\rm d}P^k_{t-s}
\end{aligned}
\end{equation*}

So we can write:
\begin{align*}
    \lambda_{i,t} = \lambda_{i,\infty} +  \boldsymbol{r_t}^\top\mathbb{K}_i\boldsymbol{r_t}
\end{align*}

where 
\setcounter{MaxMatrixCols}{20}

\begin{align*}
    \mathbb{K}_i = \begin{pmatrix} 
    K^i_{11}(1,1)  & \dots & K^i_{11}(1,q) & \dots &\frac{1}{2}K^i_{1N}(1,1) & \dots & \frac{1}{2}K^i_{1N}(1,q) \\
    \vdots & \ddots &\vdots && \vdots & \ddots  &\vdots\\
    K^i_{11}(q,1) & \dots & K^i_{11}(q,q) & \dots &\frac{1}{2}K^i_{1N}(q,1) & \dots & \frac{1}{2}K^i_{1N}(q,q)\\
    &\vdots&& \ddots&&\vdots&& \\
    \frac{1}{2}K^i_{1N}(1,1)  & \dots & \frac{1}{2}K^i_{1N}(1,q) &\dots &K^i_{NN}(1,1) & \dots & K^i_{NN}(1,q) \\
    \vdots & \ddots &\vdots & & \vdots & \ddots  &\vdots\\
    \frac{1}{2}K^i_{1N}(q,1) & \dots & \frac{1}{2}K^i_{1N}(q,q) & \dots &K^i_{NN}(q,1) & \dots & K^i_{NN}(q,q)\\
    \end{pmatrix}, \qquad \boldsymbol{r_t} = \begin{pmatrix}{\rm d}P^1_{t-1}\\\vdots\\{\rm d}P^1_{t-q}\\\vdots\\{\rm d}P^N_{t-1}\\\vdots\\{\rm d}P^N_{t-q}\end{pmatrix}
\end{align*}

So $\mathbb{K}_i$ is a symmetric blocs matrix where the bloc $j$ of the diagonal is $\text{bloc}_{jj}=\begin{pmatrix}K^i_{jj}(1,1)  & \dots & K^i_{jj}(1,q)\\
\vdots & \ddots &\vdots \\
K^i_{jj}(q,1) & \dots & K^i_{jj}(q,q)\end{pmatrix}$ and the bloc $(k,j)$ when $k< j$ is $\text{bloc}_{kj}=\frac{1}{2}\begin{pmatrix}K^i_{kj}(1,1)  & \dots & K^i_{kj}(1,q)\\
\vdots & \ddots &\vdots \\
K^i_{kj}(q,1) & \dots & K^i_{kj}(q,q)\end{pmatrix}$ (when $j<k$ the cross kernels is $K^i_{jk}$).

Thus, in this case, where $\mathbb{L}\equiv0$, sufficient condition to keep the intensity positive is to have $\mathbb{K}_i$ positive semi-definite. \newline

If we now consider the case where $\mathbb{L}\neq0$, following the same method, we can write:

\begin{align*}
    \lambda_{i,t} = \lambda_{i,\infty} + \mathbf{L}_i\boldsymbol{r_t}+ \boldsymbol{r_t}^\top\mathbb{K}_i\boldsymbol{r_t}
\end{align*}

where $\boldsymbol{r_t}$ and $\mathbb{K}_i$ are the same as before and $\mathbf{L}_i$ is 

\begin{align*}
    \mathbf{L}_i=\begin{pmatrix}L^i_{1}(1)&\dots&L^i_{1}(q)&\dots&L^i_{N}(1)&\dots&L^i_{N}(q)\end{pmatrix}^\top
\end{align*}

Assuming $\mathbb{K}_i$ is invertible, one can complete the square by writing:

\begin{align*}
    \lambda_{i,t} =&\, \lambda_{i,\infty} + \mathbf{L}_i\boldsymbol{r_t}+ \boldsymbol{r_t}^\top\mathbb{K}_i\boldsymbol{r_t}\\
    =&\, \lambda_{i,\infty} + \left(\boldsymbol{r_t} + \frac{1}{2}\mathbb{K}_i^{-1}\mathbf{L}_i\right)^\top\mathbb{K}_i\left(\boldsymbol{r_t} + \frac{1}{2}\mathbb{K}_i^{-1}\mathbf{L}_i\right) - \frac{1}{4}\mathbf{L}_i^\top\mathbb{K}_i^{-1}\mathbf{L}_i
\end{align*}




In conclusion, a sufficient condition for intensities to stay positive when $\mathbb{L}\neq0$ is to have all kernels $\mathbb{K}_i$ positive definite and  
\[
\lambda_{i,\infty} \geq \frac{1}{4}\mathbf{L}_i^\top\mathbb{K}_i^{-1}\mathbf{L}_i, \qquad \forall i
\]



\subsection{Stationary Condition in Bivariate Case }\label{app:stationaryConditionBiv}

In the bivariate case, we no longer have $\mathbb{E}({\rm d}P_u^i{\rm d}P_u^j)=0$ when $i\neq j$. Thus, when computing $\mathbb{E}(\boldsymbol{\lambda_t})$ to find the mean intensity and thus the stationary condition, we need to compute the mean of cross feedback.

We first introduce $m(dt,{\rm d}\xi^1_{},{\rm d}\xi^2)$ the joint Punctual Poisson Measure associate to the jump processes of the prices $(P^{1},P^{2})$. So it is a pure jump process with i.i.d jump sizes $(\xi^1\xi^2)$ of common law $p$ on $(\mathds{R},\mathcal{B}(\mathds{R}))$. We assume   $\int_{\mathds{R}}\xi^1\xi^2p({\rm d}\xi^1_{},{\rm d}\xi^2_{})=\psi_1\psi_2$.

\begin{align*}
     \frac{1}{\psi_1\psi_2}\mathbb{E}\Big(\int_{-\infty}^t\int_{-\infty}^tK_{12}(t-s,t-u){\rm d}P^1_s{\rm d}P^2_u\Big)  &=\frac{1}{\psi_1\psi_2}\mathbb{E}\Big(\int_{-\infty}^tK_{12}(t-u,t-u){\rm d}P^1_u{\rm d}P^2_u\Big)\\
    &= \frac{1}{\psi_1\psi_2}\mathbb{E}\Big(\int_{\mathds{R}}\int_{-\infty}^tK_{12}(t-u,t-u)s^1_us^2_u\xi^1_{u}\xi^2_{u}m({\rm d}u,{\rm d}\xi^1_{},{\rm d}\xi^2_{})\Big)\\
    &= \frac{1}{\psi_1\psi_2}\mathbb{E}\Big(\int_{\mathds{R}}\int_{-\infty}^tK_{12}(t-u,t-u)s^1_us^2_u\xi^1_{u}\xi^2_{u}\mu^{\text{c}}_u{\rm d}up({\rm d}\xi^1_{},{\rm d}\xi^2_{})\Big)\\
    &= \frac{1}{\psi_1\psi_2}\mathbb{E}\Big(\int_{-\infty}^tK_{12}(t-u,t-u)s^1_us^2_u\mu^{\text{c}}_u{\rm d}u\Big)\mathbb{E}\Big(\int_{\mathds{R}}\xi^1_{u}\xi^2_{u}p({\rm d}\xi^1_{},{\rm d}\xi^2_{})\Big)\\
    &= \int_{-\infty}^tK_{12}(t-u,t-u)\mathbb{E}(s^1_us^2_u)\mathbb{E}(\mu^{\text{c}}_u){\rm d}u\\
    &= \rho\overline{\mu^{\text{c}}}\int_{-\infty}^tK_{12}(t-u,t-u){\rm d}u
\end{align*}



\subsection{Yule-Walker Equations: Complement}\label{app:YWD}

We write the Yule-Walker equations for the matrix and one for $\mathbb{D}_{\cross}$, the calculations can be made as in Appendix 1 of \cite{blanc2017quadratic}:
\setcounter{equation}{0} 

\begin{equation}\label{eqApp:D_Yule_ind_mQH}
    \begin{aligned}
    \mathbb{D}_{\cross}(\tau_1,\tau_2)
    =& \int_{\tau_1^+}^{+\infty}\mathbb{K}_{\text{d}}(u)\mathbb{D}_{\times}(\tau_1-u,\tau_2-u){\rm d}u\\&+ 2\int_{\tau_1^+}^{+\infty}\mathbb{K}_{\text{d}}(u,\tau_1)\begin{pmatrix}\Db_{112}(u-\tau_1,\tau_2-\tau_1)&0\\0&\Db_{221}(u-\tau_1,\tau_2-\tau_1)\end{pmatrix}{\rm d}u\\
    &+\int_{\tau_1^+}^{+\infty} \mathbf{K}_{\cross}(u,\tau_1)\begin{pmatrix}0\\\Db_{211}(u-\tau_1,\tau_2-\tau_1)\end{pmatrix}^\top{\rm d}u+\int_{\tau_1^+}^{+\infty} \mathbf{K}_{\cross}(\tau_1,u)\begin{pmatrix}\Db_{122}(u-\tau_1,\tau_2-\tau_1)\\0\end{pmatrix}^\top{\rm d}u
    \end{aligned}
\end{equation}

In the bivariate case, we need to consider ${\rm d}N^c$. Thus, $\mathbb{C}$ becomes a 3$\times$3 matrix, and the three-point correlation now also considers the components $\mathbb{D}_{cij}$. 
With this in mind, we can calculate the Yule-Walker equation for $\mathbb{C}$ in the bivariate case:
\begin{align*}
    \mathbb{C}(\tau)=& \, \overline{\bblambda}\mathbb{Q}_{\text{d}}(\tau)+\int_{0}^{+\infty} \mathbb{Q}_{\text{d}}(u,u)\mathbb{C}(\tau-u){\rm d}u+2
    \int_{0}^{+\infty}\int_{u^+}^{+\infty} \mathbb{Q}_{\text{d}}(\tau+u,\tau+v)\mathbb{D}_{\text{d}}(u,v){\rm d}v{\rm d}u\\&+ \int_{0}^{+\infty}\int_{u^+}^{+\infty}\mathbf{Q}_{\cross}(\tau+u,\tau+v)\begin{pmatrix}
    \Db_{112}\\\Db_{212}\\\Db_{c12}
    \end{pmatrix}^\top(u,v){\rm d}v{\rm d}u+ \int_{0}^{+\infty}\int_{u^+}^{+\infty}\mathbf{Q}_{\cross}(\tau+u,\tau+v)\begin{pmatrix}
    \Db_{112}\\\Db_{212}\\\Db_{c12}
    \end{pmatrix}^\top(u,v){\rm d}u{\rm d}v\\
    &+\rho\int_{0}^{+\infty}\begin{pmatrix}
    Q_{12}^1\\Q_{12}^2\\Q_{12}^c
    \end{pmatrix}(u)\begin{pmatrix}
    \Cb_{c1}\\\Cb_{c2}\\\Cb_{c3}
    \end{pmatrix}^\top(\tau-u){\rm d}u
    +\int_{0}^{+\infty}\begin{pmatrix}
    Q_{11}^1\\Q_{22}^2\\Q_{cc}^c
    \end{pmatrix}(u)\begin{pmatrix}
    \Cb_{c1}\\\Cb_{c2}\\\Cb_{c3}
    \end{pmatrix}^T(\tau-u){\rm d}u
\end{align*}

where $\overline{\bblambda}$ is a $2 \times 2$ matrix defined as
\begin{align*} \overline{\bblambda}:=\begin{pmatrix}
    \Bar{\lambda}_1 & \Bar{\mu}_{\text{c}}\\
    \Bar{\mu}_{\text{c}} &  \Bar{\lambda}_2 
    \end{pmatrix}.
\end{align*}

\subsection{Asymptotic Behavior of Decaying Power Law Kernels}\label{app:assym}

\subsubsection{Asymptotic Forms}\label{app:assym_forms}

We consider decaying power law kernels, such that:
\begin{align*}
    \mathbb{K}_{\text{d}}(\tau) \underset{\tau\xrightarrow{}+\infty}{\sim} \begin{pmatrix}
    k_{d1}\tau^{-1-\epsilon_{\text{d}}}&k_{12}\tau^{-1-\epsilon_{\text{o}}}\\k_{21}\tau^{-1-\epsilon_{\text{o}}}&k_{d2}\tau^{-1-\epsilon_{\text{d}}}
    \end{pmatrix} \quad    \mathbb{K}_{\text{d}}(\tau v_1,\tau v_2) \underset{\tau\xrightarrow{}+\infty}{\sim}\begin{pmatrix}
    \Tilde{K}_{11}(v_1,v_2)\tau^{-2\rho_{\text{d}}}&\Tilde{K}_{12}(v_1,v_2)\tau^{-2\rho_{\text{o}}}\\\Tilde{K}_{21}(v_1,v_2)\tau^{-2\rho_{\text{o}}}&\Tilde{K}_{22}(v_1,v_2)\tau^{-2\rho_{\text{d}}}
    \end{pmatrix}
\end{align*}

\begin{align*}
    \mathbb{K}_{\cross}(\tau) \underset{\tau\xrightarrow{}+\infty}{\sim} \begin{pmatrix}
    k_{x1}\tau^{-1-\epsilon_{\cross}}\\k_{x2}\tau^{-1-\epsilon_{\cross}}
    \end{pmatrix}\quad \mathbb{K}_{\cross}(\tau v_1,\tau v_2) \underset{\tau\xrightarrow{}+\infty}{\sim} \begin{pmatrix}
    \Tilde{K}^1_{12}(v_1,v_2)\tau^{-2\rho_{\cross}}\\\Tilde{K}^2_{12}(v_1,v_2)\tau^{-2\rho_{\cross}}
    \end{pmatrix}
\end{align*}

Where $\Tilde{K}_{11}$, $\Tilde{K}_{12}$, $\Tilde{K}_{21}$, $\Tilde{K}_{22}$, $\Tilde{K}^1_{12}$ and $\Tilde{K}^2_{12}$ are bounded. Given the Yule-Walker equations of Section \ref{sec:YW}, we expect the correlation structure to have a similar form. So we look for them as decaying power law functions with parameters defined such as:

\begin{align*}
    \mathbb{C}(\tau) \underset{\tau\xrightarrow{}+\infty}{\sim} \begin{pmatrix}
    c_{d1}\tau^{-\beta_{\text{d}}}&c_{12}\tau^{-\beta_{\text{o}}}\\c_{21}\tau^{-\beta_{\text{o}}}&c_{d2}\tau^{-\beta_{\text{d}}}
    \end{pmatrix}\quad  \mathbb{D_{\text{d}}}(\tau v_1,\tau v_2) \underset{\tau\xrightarrow{}+\infty}{\sim} \begin{pmatrix}
    \Tilde{\Db}_{111}(v_1,v_2)\tau^{-2\delta_{\text{d}}}&\Tilde{\Db}_{122}(v_1,v_2)\tau^{-2\delta_{\text{o}}}\\\Tilde{\Db}_{211}(v_1,v_2)\tau^{-2\delta_{\text{o}}}&\Tilde{\Db}_{222}(v_1,v_2)\tau^{-2\delta_{\text{d}}}
    \end{pmatrix}
\end{align*}

\begin{align*}
    \mathbb{D_{\cross}}(\tau) \underset{\tau\xrightarrow{}+\infty}{\sim} \begin{pmatrix}
    d_{112}\tau^{-\beta^{\cross}_{\text{d}}}&d_{121}\tau^{-\beta_{\text{o}}^{\cross}}\\d_{212}\tau^{-\beta_{\text{o}}^{\cross}}&d_{221}\tau^{-\beta_{\text{d}}^{\cross}}
    \end{pmatrix}\quad   \mathbb{D_{\cross}}(\tau v_1,\tau v_2) \underset{\tau\xrightarrow{}+\infty}{\sim} \begin{pmatrix}
    \Tilde{\Db}_{112}(v_1,v_2)\tau^{-2\delta_{\cross}}&\Tilde{\Db}_{121}(v_1,v_2)\tau^{-2\delta_{\cross}}\\\Tilde{\Db}_{212}(v_1,v_2)\tau^{-2\delta_{\cross}}&\Tilde{\Db}_{221}(v_1,v_2)\tau^{-2\delta_{\cross}}
    \end{pmatrix}
\end{align*}

As in \cite{blanc2017quadratic}, we make the following hypothesis on the exponents:
$\rho_{\text{d}}>\frac{1}{2}$, $\rho_{\text{o}}>\frac{1}{2}$, $\rho_{\cross}>\frac{1}{2}$, so the first and second moments are finite. 
Moreover, we focus on the cases:
\begin{itemize}
    \item Non critical case $(n_H<1)$, where we assume $0<\epsilon_{\text{d}}<1$, $0<\epsilon_{\text{o}}<1$, 
    \item Critical case $(n_H=1)$, where we assume $0<\epsilon_{\text{d}}<\frac{1}{2}$, $0<\epsilon_o<\frac{1}{2}$,
\end{itemize}

In the non-critical case $(n<1)$, the method consists in replacing $\mathbb{K}_{\text{d}}$, $\mathbb{K}_{\cross}$, $\mathbb{C}$, $\mathbb{D_{\text{d}}}$ and $\mathbb{D_{\cross}}$ by their asymptotic expressions presented above in the Yule-Walker equations (Eq.~\eqref{eq:C_YuleWalker_mQH}, Eq.~\eqref{eq:D_Yule_ind_mQH} and Eq.~\eqref{eqApp:D_Yule_ind_mQH}) and study the limit $\tau \xrightarrow{}+\infty$. \newline

In the following, we describe the method to find the relationship between the exponents of the correlation structures and the exponents of the kernels in the critical case. In a next section, we give the resulting relationships for both the critical and non critical case.

\subsubsection{Method for the Asymptotic Study in the Critical Case of QHawkes process}\label{app:assym_met}

In critical case, when $n_H=1$, the relationship between auto-correlation structures exponents and kernel exponents can not be determined by looking at the limit $\tau \rightarrow +\infty$. Thus, to overcome this difficulty, we use a second method to investigate asymptotic behavior using Fourier-domain. \newline

This method for the linear Hawkes can be found in \cite{hawkes1971point}. We present it now for the quadratic Hawkes process. For the sake of simplicity we limit ourselves to the 1D case here, the multivariate case can be worked out similarly.\newline

The definitions of the autocorrelation structures in the 1D case can be found in \cite{blanc2017quadratic} and are substantially similar to those used here. 

We define the Fourier transform of a function $f$ such as: 

\begin{equation*}
    \hat{f}(\omega) = \int_\mathbb{R} f(t)e^{-i\omega t} {\rm d}t
\end{equation*}

\textbf{Step 1: Find the regularity of the Yule-Walker terms:}

As for the linear case, we start from the Yule-Walker equation on $\Cb$ for $\tau\neq0$ (see equation (9) in \cite{blanc2017quadratic}): 
\begin{equation}\label{eq:appQHFourier}
    \Cb(\tau) 
    = K(\tau)\bar{\lambda} + \int_{0}^{+\infty}K(\tau-u)\Cb(u){\rm d}u + 2\int_{0+}^{+\infty}\int_{u+}^{+\infty}K(\tau+u,\tau+r)\Db(u,r){\rm d}r{\rm d}u,
\end{equation}
and its extension in 0 in Fourier domain gives:
\begin{equation}\label{eq:C_extended_Fourier}
    \hat{\Cb}^*(\omega) = \bar{\lambda}+\hat{\Cb}(\omega).
\end{equation}

To use the Fourier transform, we need to extend $K$ and $\Db$ for $\tau,\tau_1,\tau_2<0$. Thus, we consider the function $K$ defined on $\mathbb{R}$ with $K(\tau)=0$ for $\tau<0$ similarly for $\Db$. Thus, we expect to have  $\hat{K}(\omega)$ and $\hat{\Db}(\omega)$ regular in the half plan $\text{Im}(\omega)<0$.

The 2 first terms of Eq.~\eqref{eq:appQHFourier} are the same as in the Yule-Walker equation for the linear case. Hence, the regularity arguments are  the same as in \cite{hawkes1971point}.

However, we need to study the last term of Eq.~\eqref{eq:appQHFourier}, $C_3(\tau)=2\int_{0+}^{+\infty}\int_{u+}^{+\infty}K(\tau+u,\tau+r)\Db(u,r){\rm d}r{\rm d}u$. \\

In order to study the regularity we decompose $\omega$ into $\omega = \omega_R + i\omega_I$. 
Then, switching integration order and using Chasles relation we obtain: 

\begin{align*}
    \hat{C}_3(\omega) 
    =&2\int_0^{+\infty}\int_{u^+}^{+\infty}\int_{-\infty}^{-u}e^{-i\omega_R\tau}e^{\omega_I\tau}K(\tau+u,\tau+r)\Db(u,r)d\tau {\rm d}r{\rm d}u&\Big\}\quad \hat{C}_{3_1}(\omega)\\
    &+2\int_0^{+\infty}\int_{u^+}^{+\infty}\int_{-u}^{0}e^{-i\omega_R\tau}e^{\omega_I\tau}K(\tau+u,\tau+r)\Db(u,r)d\tau {\rm d}r{\rm d}u&\Big\}\quad \hat{C}_{3_2}(\omega)\\
    &+2\int_0^{+\infty}\int_{u^+}^{+\infty}\int_{0}^{+\infty}e^{-i\omega_R\tau}e^{\omega_I\tau}K(\tau+u,\tau+r)\Db(u,r)d\tau {\rm d}r{\rm d}u&\Big\}\quad \hat{C}_{3_3}(\omega)\\
\end{align*}

We study each term $\hat{C}_{3_1}$, $\hat{C}_{3_2}$ and $\hat{C}_{3_3}$  separately. 
We first have:
\begin{align*}
    \hat{C}_{3_1}(\omega) = 2\int_0^{+\infty}\int_{u^+}^{+\infty}\int_{-\infty}^{-u}e^{-i\omega_R\tau}e^{\omega_I\tau}K(\tau+u,\tau+r)\Db(u,r){\rm d}\tau {\rm d}r{\rm d}u
\end{align*}

Here $\tau < -u$ so $\tau+u<0$, so $K(\tau+u,\tau+r)=0$, so $\hat{C}_{3_1}(\omega)=0$. \\
\newline

For the second term, we switch integration order, and we obtain:

\begin{align*}
    \hat{C}_{3_2}(\omega) 
    &=2\int_{-\infty}^{0}\int_{-\tau}^{+\infty}\int_{u^+}^{+\infty}e^{-i\omega_R\tau}e^{\omega_I\tau}K(\tau+u,\tau+r)\Db(u,r) {\rm d}r{\rm d}u{\rm d}\tau
\end{align*}

In this case, we have $\tau\geq-u$, so $\tau+u\geq0$, and $r\geq u^+ \geq -\tau$, so $K(\tau+u,\tau+r)>0$. We are only interested in $\tau \xrightarrow{} -\infty$, so $\hat{C}_{3_2}(\omega)$ is regular in the half plan $\omega_I>0$. \\

For $\hat{C_{3_3}}(\omega)$, we have $\tau+u>0$ and $\tau+r>0$, so $K(\tau+u,\tau+r)\geq 0$. Since we are only worried about $\tau \xrightarrow{} +\infty$. $\hat{C_{3_3}}(\omega)$ is regular in half plan $\omega_I<0$.\\

With those results, we now introduce the Yule-Walker equation \ref{eq:appQHFourier} in Fourier domain and,  as in \cite{hawkes1971point}, the function $\hat{B}$:
\begin{align}
\begin{cases}
        \hat{\Cb}(\omega)&=\hat{K}(\omega)D + \hat{K}(\omega)\hat{\Cb}(\omega) + \hat{C_{3_2}}(\omega)+ \hat{C_{3_3}}(\omega)\\
    \hat{B}(\omega) &= \hat{K}(\omega)D + \hat{K}(\omega)\hat{\Cb}(\omega)+ \hat{C_{3_2}}(\omega)+ \hat{C_{3_3}}(\omega)-\hat{\Cb^*}(\omega)
\end{cases}
\end{align}

Thus, 
\begin{equation}\label{eqApp:mutemp1}
        \hat{\Cb}(\omega) = (1-\hat{K}(\omega))^{-1}(\hat{K}(\omega)D-\hat{B}(\omega)+ \hat{C_{3_2}}(\omega)+ \hat{C_{3_3}}(\omega))
\end{equation}

Since $\Cb$ is even, we have ($\hat{\Cb}^\top(\omega) = \hat{\Cb}(-\omega)$) and Eq.~\eqref{eqApp:mutemp1} becomes:
 \begin{align*}
     (\hat{K}(\omega)D  + \hat{C_{3_2}}(\omega)+ \hat{C_{3_3}}(\omega)  - \hat{B}(\omega))^\top&(1-\hat{K}^\top(\omega))^{-1} = (1-\hat{K}(-\omega))^{-1}(\hat{K}(-\omega)D  + \hat{C_{3_2}}(-\omega)+ \hat{C_{3_3}}(-\omega)  - \hat{B}(-\omega))\\
\end{align*}

Multiplying by $1-\hat{K}^\top(\omega)$ on the right side and by $1-\hat{K}(-\omega)$ on the left side, we then develop the expression and mark the regularity propriety below each term. 

\begin{align*}
&\underbrace{D\hat{K}^\top(\omega)}_{\text{Im}(\omega)<0} +\underbrace{\hat{C_{3_2}}^\top(\omega)}_{\text{Im}(\omega)>0}+ \underbrace{\hat{C_{3_3}}^\top(\omega)}_{\text{Im}(\omega)<0}- \underbrace{\hat{B}^\top(\omega)}_{\text{Im}(\omega)>0} -\underbrace{\hat{K}(-\omega)\hat{C_{3_2}}^\top(\omega)}_{\text{Im}(\omega)>0}-\hat{K}(-\omega)\hat{C_{3_3}}^\top(\omega)+\underbrace{\hat{K}(-\omega) \hat{B}^\top(\omega)}_{\text{Im}(\omega)>0}=\\ &\underbrace{\hat{K}(-\omega)D}_{\text{Im}(\omega)>0}  + \underbrace{\hat{C_{3_2}}(-\omega)}_{\text{Im}(\omega)<0}+ \underbrace{\hat{C_{3_3}}(-\omega)}_{\text{Im}(\omega)>0}  - \underbrace{\hat{B}(-\omega)}_{\text{Im}(\omega)<0}        -\underbrace{\hat{C_{3_2}}(-\omega)\hat{K}^\top(\omega)}_{\text{Im}(\omega)<0}- \hat{C_{3_3}}(-\omega)\hat{K}^\top(\omega) +\underbrace{\hat{B}(-\omega)\hat{K}^\top(\omega)}_{\text{Im}(\omega)<0}
\end{align*}
 
We notice we have a problem with $\hat{C_{3_3}}(-\omega)\hat{K}^\top(\omega)$ and $\hat{K}(-\omega)\hat{C_{3_3}}^\top(\omega)$. In fact, for each of them one factor is regular for $\text{Im}(\omega)>0$ and the other one for $\text{Im}(\omega)<0$... 
If we still reorder the terms to have the left side of the equality regular in the half plan $\text{Im}(\omega)<0$ and the right side of the equality regular in the half plan $\text{Im}(\omega)>0$, without defining the regularity plan for the two problematic terms, we obtain the following:

 \begin{equation}\label{eqApp:YWtemp2}
 \begin{aligned}
     &\underbrace{D\hat{K}^\top(\omega) + \hat{C_{3_3}}^\top(\omega) - \hat{C_{3_2}}(-\omega) + \hat{B}(-\omega) + \hat{C_{3_2}}(-\omega)\hat{K}^\top(\omega) - \hat{B}(-\omega)\hat{K}^\top(\omega)}_{\text{Im}(\omega)<0} -\hat{K}(-\omega)\hat{C_{3_3}}^\top(\omega)=\\
     &\underbrace{\hat{K}(-\omega)D+\hat{C_{3_3}}(-\omega)-\hat{C_{3_2}}^\top(\omega) + \hat{B}^\top(\omega) + \hat{K}(-\omega)\hat{C_{3_2}}^\top(\omega)-\hat{K}(-\omega) \hat{B}^\top(\omega) }_{\text{Im}(\omega)>0} - \hat{C_{3_3}}(-\omega)\hat{K}^\top(\omega)
 \end{aligned}
 \end{equation}
 
We now need to study in details the regularity of the problematic terms. \\

\textbf{Step 2: Expand $\hat{C_{3_3}}(-\omega)\hat{K}^\top(\omega)$ and $\hat{K}(-\omega)\hat{C_{3_3}}^\top(\omega)$}\\


We have 
\begin{align*}
    \hat{K}(-\omega)\hat{C_{3_3}}^\top(\omega)
    =& 2\int_0^{+\infty}\int_{u+}^{+\infty}\int_0^{+\infty} \Big(\hat{K}(-\omega)e^{-i\omega\tau}\Big)\Db^\top(u,r)K^\top(\tau+u,\tau+r) {\rm d}\tau {\rm d}r{\rm d}u
\end{align*}

and

\begin{align*}
    \hat{C_{3_3}}(-\omega)\hat{K}^\top(\omega)
    =&2\int_0^{+\infty}\int_{u+}^{+\infty}\int_0^{+\infty} K(\tau+u,\tau+r)\Db(u,r) \Big(e^{i\omega\tau}\hat{K}^\top(\omega)\Big){\rm d}\tau {\rm d}r{\rm d}u
\end{align*}

We look at $\hat{K}(-\omega)e^{-i\omega\tau}$ with the change of variables $\phi(t)=t+\tau$ and the Chasles relation:



\begin{align}
    \hat{K}(-\omega)e^{-i\omega\tau} = \int_0^{+\infty}\int_{-\infty}^{+\infty}e^{i\omega t}K(t+\tau,t+\tau)dtd\tau =& \int_0^{+\infty}\int_{-\infty}^{-\tau}e^{i\omega t}K(t+\tau,t+\tau){\rm d}t{\rm d}\tau\label{eq:aa}\\
    &+\int_0^{+\infty}\int_{-\tau}^{0}e^{i\omega t}K(t+\tau,t+\tau){\rm d}t{\rm d}\tau\label{eq:bb}\\
    &+\int_0^{+\infty}\int_{0}^{+\infty}e^{i\omega t}K(t+\tau,t+\tau){\rm d}t{\rm d}\tau\label{eq:cc}
\end{align}

The first term, \eqref{eq:aa}, $\int_0^{+\infty}\int_{-\infty}^{-\tau}e^{i\omega t}K(t+\tau,t+\tau){\rm d}t{\rm d}\tau$ is null, because $K(t+\tau)$ is null for $t\leq-\tau$. 
If we switch the integration order in the second term, \eqref{eq:bb}, we have $\int_0^{+\infty}\int_{-\tau}^{0}e^{i\omega t}K(t+\tau,t+\tau){\rm d}t{\rm d}\tau = \int_{-\infty}^0\int_{-t}^{+\infty}e^{i\omega t}K(t+\tau,t+\tau){\rm d}\tau {\rm d}t$ and then, we need to have a convergent exponential $e^{i\omega t}$ when $t\xrightarrow{}-\infty$, so the second term is regular in half plan $\text{Im}(\omega)<0$. And the last term, \eqref{eq:cc}, $\int_0^{+\infty}\int_{0}^{+\infty}e^{i\omega t}K(t+\tau,t+\tau){\rm d}t{\rm d}\tau$ is regular in half plan $\text{Im}(\omega)>0$.\\

We now need to consider the term $e^{i\omega\tau}\hat{K}^\top(\omega)$, similarly, we have: 
\begin{align}
    e^{i\omega\tau}\hat{K}^\top(\omega) = \int_0^{+\infty}\int_{-\infty}^{+\infty}e^{-i\omega t}K^\top(t+\tau,t+\tau){\rm d}t{\rm d}\tau =& \int_0^{+\infty}\int_{-\infty}^{-\tau}e^{-i\omega t}K^\top(t+\tau,t+\tau){\rm d}t{\rm d}\tau\label{eq:a1}\\
    &+\int_0^{+\infty}\int_{-\tau}^{0}e^{-i\omega t}K^\top(t+\tau,t+\tau){\rm d}t{\rm d}\tau\label{eq:a2}\\
    &+\int_0^{+\infty}\int_{0}^{+\infty}e^{-i\omega t}K^\top(t+\tau,t+\tau){\rm d}t{\rm d}\tau\label{eq:a3}
\end{align}

For same reasons as before, first term, \eqref{eq:a1}, is null, second term, \eqref{eq:a2}, is regular in half plan $\text{Im}(\omega)>0$, and last term, \eqref{eq:a3}, is regular in half plan $\text{Im}(\omega)<0$. \\

Wrapping up those results, we have
\begin{align*}
    \hat{K}(-\omega) \hat{C}_{3_3}^\top(\omega) 
    =&2\int_0^{+\infty}\int_{u+}^{+\infty}\int_0^{+\infty} \int_{-\tau}^{0}e^{i\omega t}K(t+\tau,t+\tau){\rm d}t \Db^\top(u,r)K^\top(\tau+u,\tau+r){\rm d}\tau {\rm d}r{\rm d}u &\Big\}\quad F_1(\omega)\\
    &+2\int_0^{+\infty}\int_{u+}^{+\infty}\int_0^{+\infty} \int_{0}^{+\infty}e^{i\omega t}K(t+\tau,t+\tau){\rm d}t \Db^\top(u,r)K^\top(\tau+u,\tau+r){\rm d}\tau {\rm d}r{\rm d}u&\Big\}\quad F_2(\omega)\\
    =& F_1(\omega)+ F_2(\omega)
\end{align*}
with $F_1(\omega)$ being regular for $\text{Im}(\omega)< 0$ and $F_2(\omega)$ for $\text{Im}(\omega) >0$. 

Similarly, 
\begin{align*}
    \hat{C_{3_3}}(-\omega)\hat{K}^\top(\omega) 
    =&2\int_0^{+\infty}\int_{u+}^{+\infty}\int_0^{+\infty} K(\tau+u,\tau+r)\Db(u,r)\int_{-\tau}^{0}e^{-i\omega t}K^\top(t+\tau,t+\tau){\rm d}t{\rm d}\tau {\rm d}r{\rm d}u&\Big\}\, E_1(-\omega)\\
    &+2\int_0^{+\infty}\int_{u+}^{+\infty}\int_0^{+\infty} K(\tau+u,\tau+r)\Db(u,r)\int_{0}^{+\infty}e^{-i\omega t}K^\top(t+\tau,t+\tau){\rm d}t{\rm d}\tau {\rm d}r{\rm d}u&\Big\}\, E_2(-\omega)\\
    =& E_1(-\omega)+ E_2(-\omega)
\end{align*}
with $E_1(-\omega)$ being regular for $\text{Im}(\omega)> 0$ and $E_2(-\omega)$ for $\text{Im}(\omega) <0$.\\

\textbf{Step 3: Wrap up}\\

We can now go back to Eq.~\eqref{eqApp:YWtemp2} which becomes: 

 
 \begin{align*}
     &D\hat{K}^\top(\omega) + \hat{C_{3_3}}^\top(\omega) - \hat{C_{3_2}}(-\omega) + \hat{B}(-\omega) + \hat{C_{3_2}}(-\omega)\hat{K}^\top(\omega) - \hat{B}(-\omega)\hat{K}^\top(\omega) -F_1(\omega)+E_2(-\omega)=\\
     &\hat{K}(-\omega)D+\hat{C_{3_3}}(-\omega)-\hat{C_{3_2}}^\top(\omega) + \hat{B}^\top(\omega) + \hat{K}(-\omega)\hat{C_{3_2}}^\top(\omega)-\hat{K}(-\omega) \hat{B}^\top(\omega)  -E_1(-\omega)+F_2(\omega)
 \end{align*}
 
 We then have the left side regular in half plan $\text{Im}(\omega)<0$ and the right one in half plan $\text{Im}(\omega)>0$. So as in \cite{hawkes1971point}, we can say that the lower side is null, and obtain an expression for $\hat{B}$: 
 \begin{align*}
     &0=\hat{K}(-\omega)D+\hat{C_{3_3}}(-\omega)-\hat{C_{3_2}}^\top(\omega) + \hat{B}^\top(\omega) + \hat{K}(-\omega)\hat{C_{3_2}}^\top(\omega)-\hat{K}(-\omega) \hat{B}^\top(\omega)  -E_1(-\omega)+F_2(\omega)\\
     &\Longrightarrow\begin{cases}
          \hat{B}^\top(\omega)&=-\Big(\hat{K}(-\omega)D+\hat{C_{3_3}}(-\omega)-\hat{C_{3_2}}^\top(\omega)  + \hat{K}(-\omega)\hat{C_{3_2}}^\top(\omega)  -E_1(-\omega)+F_2(\omega)\Big)^\top(1-\hat{K}^\top(-\omega))^{-1}\\
      \hat{B}(\omega)&=-\Big(D\hat{K}^\top(-\omega)+\hat{C}_{3_3}^\top(-\omega)-\hat{C}_{3_2}(\omega)  + \hat{C}_{3_2}(\omega) \hat{K}^\top(-\omega) -E_1^\top(-\omega)+F_2^\top(\omega)\Big)(1-\hat{K}^\top(-\omega))^{-1} 
     \end{cases}
 \end{align*}

Then we inject the expression of $\hat{B}$ in $\hat{\Cb^*}$, and after some manipulations where we use $\hat{C_{3_3}}(\omega)\hat{K}^\top(-\omega)=E_1(\omega) + E_2(\omega)$: 
 
\begin{align*}
    \hat{\Cb}(\omega)
    =&(1-\hat{K}(\omega))^{-1}\Big(\hat{K}(\omega)D + \hat{C_{3_3}}(\omega)-\hat{K}(\omega)D\hat{K}^\top(-\omega) -  E_1(\omega) - E_2(\omega)\\&+
    D\hat{K}^\top(-\omega)+\hat{C}_{3_3}^\top(-\omega)   -E_1^\top(-\omega)+F_2^\top(\omega)
    \Big)(1-\hat{K}^\top(-\omega))^{-1}.
\end{align*}

Note that in 1D, the transpose and commutative operations do not really matter. However, in order to use this method in the multivariate case we presented the method paying attention to it. 


In 1D, we have $F_1(\omega) = E_1(\omega)$ and $F_2(\omega) = E_2(\omega)$, the transpose are equal the element itself, and the products are commutative, thus:
\begin{align*}
    \hat{\Cb}(\omega) 
    =& (1-\hat{K}(\omega))^{-1}\Big(\hat{K}(\omega)D + \hat{C_{3_3}}(\omega)-\hat{K}(\omega)D\hat{K}(-\omega) -  E_1(\omega) +
    \hat{K}(-\omega)D+\hat{C_{3_3}}(-\omega)   -E_1(-\omega)
    \Big)(1-\hat{K}(-\omega))^{-1}.
\end{align*}

Finally, using Eq.~\eqref{eq:C_extended_Fourier}, we obtain for QHawkes in 1D:

\begin{equation*}
    \hat{\Cb}^\star(\omega) = (1-\hat{K}(\omega))^{-1}\Big( D + \hat{C_{3_3}}(\omega)+\hat{C_{3_3}}(-\omega) -  E_1(\omega) -E_1(-\omega)\Big)(1-\hat{K}(-\omega))^{-1}.
\end{equation*}

For the multivariate case the matrices we integrate in $C_{3_3}$, $E_1$,  $E_2$, $F_1$ and $F_2$, will be a mix of $\mathbb{D}_{\text{d}}$, $\mathbb{D}_{\cross}$, $\mathbb{K}_{\text{d}}$ and $\mathbf{K}_{\cross}$. 



\subsection{Asymptotic Study- Results Tables}\label{app:resultsTab}

Based on the notations in Appendix \ref{app:assym_forms}, we give here the results of the asymptotic behavior of covariance structures when considering power law decaying kernels. 

In both non critical and critical case, we have the same result for the exponent of non time diagonal parts $\delta_{\cross}=\rho_{\cross}+\delta_o-\frac{1}{2}$ and $\delta_{\text{o}}=\rho_{\text{o}}$. The exponent $\delta_{\text{d}}=\rho_{\text{d}}$ can take 2 values:

\begin{enumerate}
    \item  $\delta_{\text{d}}=\rho_{\text{d}}$
    \item $\delta_{\text{d}}=2\rho_{\cross}+\delta_{\text{o}}-\frac{1}{2}$
\end{enumerate}

From empirical observations, we see that it makes sense if the diagonal terms persist longer in time than the non diagonal terms. Hence, we expect $\delta_{\text{d}}<\delta_{\text{o}}$. Considering this hypothesis, we would only keep the first case, $\delta_{\text{d}}=\rho_{\text{d}}<\rho_{\text{o}}$.

For the time diagonal exponent ($\beta_{\text{d}}$,$\beta_{\text{o}}$,$\beta_{\text{d}}^{\cross}$, $\beta_{\text{o}}^{\cross}$), we need to make a difference between the non critical case ($n_H<1$), and the critical case ($n_H=1$). We call $\epsilon=\min(\epsilon_{\text{d}},\epsilon_{\text{o}})$ and $\rho=\frac{1}{2}\min(\delta_{\text{d}}+\rho_{\text{d}},\delta_{\text{o}}+\rho_{\text{o}},\delta_{\text{d}}+\rho_{\text{o}},\delta_{\text{o}}+\rho_{\text{d}},\delta_{\cross}+\rho_{\cross})=\frac{1}{2}\min(\rho_{\text{d}}+\rho_{\text{o}},2\rho_{\text{o}},2\rho_{\text{2}},\delta_{\cross}+\rho_{\cross})$.

\begin{table}[h]
\centering
{\bgroup
\def\arraystretch{1.5}
\begin{tabular}{|c||c|c|}
\hline
\multicolumn{3}{|c|}{Non Critical}\\
\hhline{|===|}
   & if $\frac{3+\epsilon_{\text{d}}}{4}<\min(\rho_{\text{d}},\rho_{\cross}+\frac{\rho_{\text{o}}}{2}+\frac{1}{4})$ & if $\frac{3+\epsilon_{\text{d}}}{4}>\min(\rho_{\text{d}},\rho_{\cross}+\frac{\rho_{\text{o}}}{2}+\frac{1}{4})$\\\hline
$\beta_{\text{d}}$  & $\beta_{\text{d}}=1+\epsilon_{\text{d}}$  & $\beta_{\text{d}}=\min(4\rho_{\cross}+2\rho_{\text{o}}-1,4\rho_{\text{d}}-2)$\\
 \hline
   & if $\frac{3+\epsilon_{\text{o}}}{4}<\min(\frac{\rho_{\text{d}}+\rho_{\text{o}}}{2},\rho_{\cross}+\frac{\rho_{\text{o}}}{2}+\frac{1}{4})$ & if $\frac{3+\epsilon_{\text{o}}}{4}>\min(\frac{\rho_{\text{d}}+\rho_{\text{o}}}{2},\rho_{\cross}+\frac{\rho_{\text{o}}}{2}+\frac{1}{4})$\\\hline
$\beta_{\text{o}}$  & $\beta_{\text{o}}=1+\epsilon_{\text{o}}$  & $\beta_{\text{o}}=\min(4\rho_{\cross}+2\rho_{\text{o}}-1,2(\rho_{\text{d}}+\rho_{\text{o}})-2)$\\ \hhline{|===|}
\multicolumn{3}{|c|}{Critical}\\
\hhline{|===|}
   & if $\rho<\frac{3}{2}$  & if $\rho>\frac{3}{2}$\\\hline
$\beta_{\text{d}}=\beta_{\text{o}}=\beta$ &$\beta=4\rho-2-2\epsilon$  & $\beta=1-2\epsilon$    \\
 \hline
\end{tabular}
\egroup}
\caption{$\mathbb{C}$ exponents for both non critical and critical case, when using the following notations $\epsilon=\min(\epsilon_{\text{d}},\epsilon_{\text{o}})$ and $\rho=\frac{1}{2}\min(\delta_{\text{d}}+\rho_{\text{d}},\delta_{\text{o}}+\rho_{\text{o}},\delta_{\text{d}}+\rho_{\text{o}},\delta_{\text{o}}+\rho_{\text{d}},\delta_{\cross}+\rho_{\cross})$}
\label{tab: Cri_C}
\end{table}

\begin{table}[h]
\centering
{\bgroup
\def\arraystretch{1.5}
\begin{tabular}{|c||c|c|}
\hline
\multicolumn{3}{|c|}{Non Critical}\\
\hhline{|===|}
  & if $\rho_{\cross}<2\rho_{\text{d}}-\frac{1}{2}$ & if $\rho_{\cross}>2\rho_{\text{d}}-\frac{1}{2}$\\\hline
$\beta_{\text{d}}^{\cross}$  & $\beta_{\text{d}}^{\cross}=2\rho_{\cross}+\rho_{\text{o}}-1$  & $\beta_{\text{d}}^{\cross}=2\rho_{\text{d}}+\delta_{\cross}-1$\\
 \hline
 & if $\rho_{\cross}<2\rho_{\text{o}}-\frac{1}{2}$ & if $\rho_{\cross}>2\rho_{\text{d}}-\frac{1}{2}$\\\hline
$\beta_{\text{o}}^{\cross}$ &$\beta_{\text{o}}^{\cross}=2\rho_{\cross}+\rho_{\text{o}}-1$  & $\beta_{\text{o}}^{\cross}=2\rho_{\text{o}}+\delta_{\cross}-1$  \\ \hhline{|===|}
\multicolumn{3}{|c|}{Critical}\\
\hhline{|===|}
  & if $\frac{2\rho_{\cross}+\delta_{o}}{3}>\frac{2}{3}$ and  $\frac{2\rho_d+\delta_{\times}}{3}>\frac{2}{3}$& else\\\hline
$\beta_{\text{d}}^{\cross}$ &$\beta_{\text{d}}^{\cross}=1-\epsilon$  & $\beta_{\text{d}}^{\cross}=\min(2\rho_{\text{d}}+\delta_{\cross}-\epsilon,2\rho_{\cross}+\delta_{\text{o}}-\epsilon)$    \\
 \hline
 &if $\frac{2\rho_{\cross}+\delta_{o}}{3}>\frac{2}{3}$ and $\frac{2\rho_o+\delta_{\times}}{3}>\frac{2}{3}$ & else\\\hline
$\beta_{\text{o}}^{\cross}$&$\beta_{\text{o}}^{\cross}=1-\epsilon$  &  $\beta_{\text{o}}^{\cross}=\min(2\rho_{\text{o}}+\delta_{\cross}-\epsilon,2\rho_{\cross}+\delta_{\text{o}}-\epsilon)$  \\ \hline
\end{tabular}
\egroup}
\caption{$\mathbb{D}_{\cross}$ "time diagonal" exponents for both non critical and critical case, when using the following notations $\epsilon=\min(\epsilon_{\text{d}},\epsilon_{\text{o}})$ and $\rho=\frac{1}{2}\min(\delta_{\text{d}}+\rho_{\text{d}},\delta_{\text{o}}+\rho_{\text{o}},\delta_{\text{d}}+\rho_{\text{o}},\delta_{\text{o}}+\rho_{\text{d}},\delta_{\cross}+\rho_{\cross})$}
\label{tab: Cri_C}
\end{table}

\newpage
\subsection{Computation of the Infinitesimal Generator}\label{app:infiniV2}

We consider the processes, introducing a time scale T > 0 that will eventually diverge, we
define the processes $\bar{h}^{T}_1(t)= h_1(tT)$ ,$\bar{h}^{T}_2(t)= h_2(tT)$, $\bar{z}^{T}_1(t)= z_1(tT)$ and $\bar{z}^{T}_2(t)= z_2(tT)$, $\bar{N}^{1T}_t= N^1_{tT}$, $\bar{N}^{2T}_t= N^2_{tT}$ with the parameters $(\beta_{1T},\beta_{2T})$ and  $(\omega_{1T},\omega_{2T})$ that may depend on
$T$. In the decaying exponential case, we have

\begin{align*}
    \begin{cases}
    d(h_1)_t &= \beta_{1}(-(h_1)_t{\rm d}t + {\rm d}N^1_t)\\
    d(h_2)_t &= \beta_{2}(-(h_2)_t{\rm d}t + {\rm d}N^2_t)\\
    d(z_1)_t &= -\omega_{1}(z_1)_t{\rm d}t + \omega_{1}{\rm d}P^{1}_t\\
    d(z_2)_t &= -\omega_{2}(z_2)_t{\rm d}t + \omega_{2}{\rm d}P^{2}_t
    \end{cases} \Longrightarrow \begin{cases}
    d(\bar{h}^{T}_1)_t &= -\beta_{1T}((\bar{h}^{T}_1)_tT{\rm d}t + n_H{\rm d}N^{1T}_t)\\
    d(\bar{h}^{T}_2)_t &= -\beta_{2T}((\bar{h}^{T}_2)_tT{\rm d}t + n_H{\rm d}N^{2T}_t)\\
    d(\bar{z}^{T}_1)_t &= -\omega_{1T}(\bar{z}^{T}_1)_tT{\rm d}t + \omega_{1T}{\rm d}P^{1T}_t\\
    d(\bar{z}^{T}_2)_t &= -\omega_{2T}(\bar{z}^{T}_2)_tT{\rm d}t + \omega_{2T}{\rm d}P^{2T}_t
    \end{cases}
\end{align*}



The intensity of, respectively, the first and second process are given by $T[\lambda^1_\infty+n^1_{H,1}\bar{h}^{T}_1 + n^1_{H,2}\bar{h}^{T}_2 + (a^1_{Z,1}\bar{z}^{T}_1)^2 + (a^1_{Z,2}\bar{z}^{T}_2)^2 +a^1_{Z,\cross}\bar{z}^{T}_1\bar{z}^{T}_2]$ and $T[\lambda^2_\infty+n^2_{H,1}\bar{h}^{T}_1 + n^2_{H,2}\bar{h}^{T}_2 + (a^2_{Z,1}\bar{z}^{T}_1)^2 + (a^2_{Z,2}\bar{z}^{T}_2)^2 +a^2_{Z,\cross}\bar{z}^{T}_1\bar{z}^{T}_2]$. Moreover, the price processes can either go up or down with same probability $\frac{1}{2}$.

Thus, the infinitesimal generator can be written using \textit{Theorem 1.22 of Oksendal and Sulem}:

\begin{align*}
    Af^T(h_1,h_2,z_{1},z_{2}) =& -T\beta_{1T}h_1\partial_{h_1}f
 -T\beta_{2T}h_2\partial_{h_2}f 
 -\omega_{1T}Tz_{1}\partial_{z_{1}}f 
 -\omega_{2T}Tz_{2}\partial_{z_{2}}f \\
 &+ T[\lambda^1_\infty+n^1_{H,1}h_1 + n^1_{H,2}h_2 + (a^1_{Z,1}z_1)^2 + (a^1_{Z,2}z_2)^2 +a^1_{Z,\cross}z_{2}z_1]\Big(\\
&\frac{1}{2}f(h_1+ \beta_{T1},h_2,z_{1} + \omega_{1T},z_{2} )+\frac{1}{2}f(h_1+ \beta_{T1},h_2,z_{1} - \omega_{1T},z_{2})-f(h_1,h_2,z_1,z_2)\Big)\\
&+T[\lambda^2_\infty+n^2_{H,1}h_1 + n^2_{H,2}h_2 + (a^2_{Z,1}z_1)^2 + (a^2_{Z,2}z_2)^2 +a^2_{Z,\cross}z_{2}z_1]\Big(\\
 &\frac{1}{2}f(h_1,h_2+ \beta_{T2},z_{1},z_{2} -\omega_{2T})+\frac{1}{2}f(h_1,h_2+ \beta_{T2},z_{1} ,z_{2} +\omega_{2T})-f(h_1,h_2,z_1,z_2)\Big)
\end{align*}

As in \cite{blanc2017quadratic}, we use the scaling $\beta_{1T} = \frac{\bar{\beta}_1}{T}$, $\beta_{2T} = \frac{\bar{\beta}_2}{T}$, $\omega_{1T} = \frac{\bar{\omega}_1}{T}$, $\omega_{2T} = \frac{\bar{\omega}_2}{T}$ and the Taylor development when $T\rightarrow+\infty$ which result in: 
\begin{align*}
    Af^T(h_1,h_2,z_{1},z_{2}) =& -\bar{\beta}_{1}h_1\partial_{h_1}f
 -\bar{\beta}_{2}h_2\partial_{h_2}f 
 -\bar{\omega}_{1}z_{1}\partial_{z_{1}}f 
 -\bar{\omega}_{2}z_{2}\partial_{z_{2}}f \\&
+ T[\lambda^1_\infty+n^1_{H,1}h_1 + n^1_{H,2}h_2 + (a^1_{Z,1}z_1)^2 + (a^1_{Z,2}z_2)^2 +a^1_{Z,\cross}z_{2}z_1]\Big(
\frac{\bar{\beta}_1}{T}\partial_{h_1}f+
\frac{(\bar{\omega}_1)^2}{2T}\partial^2_{z_1z_1}f+o(\frac{1}{T})\Big)\\&
+ T[\lambda^2_\infty+n^2_{H,1}h_1 + n^2_{H,2}h_2 + (a^2_{Z,1}z_1)^2 + (a^2_{Z,2}z_2)^2 +a^2_{Z,\cross}z_{2}z_1]\Big(
\frac{\bar{\beta}_2}{T}\partial_{h_2}f+\frac{(\bar{\omega}_2)^2}{2T}\partial^2_{z_2z_2}f+o(\frac{1}{T})\Big)
\end{align*}

Thus, the infinitesimal generator is given by:
\begin{align*}
    Af^\infty(h_1,h_2,z_{1},z_{2}) =& -\bar{\beta}_{1}h_1\partial_{h_1}f
 -\bar{\beta}_{2}h_2\partial_{h_2}f 
 -\bar{\omega}_{1}z_{1}\partial_{z_{1}}f 
 -\bar{\omega}_{2}z_{2}\partial_{z_{2}}f \\&
+ [\lambda^1_\infty+n^1_{H,1}h_1 + n^1_{H,2}h_2 + (a^1_{Z,1}z_1)^2 + (a^1_{Z,2}z_2)^2 +a^1_{Z,\cross}z_{2}z_1]\Big(
n_H\bar{\beta}_1\partial_{h_1}f+
\frac{(\bar{\omega}_1)^2}{2}\partial^2_{z_1z_1}f\Big)\\&
+ [\lambda^2_\infty+n^2_{H,1}h_1 + n^2_{H,2}h_2 + (a^2_{Z,1}z_1)^2 + (a^2_{Z,2}z_2)^2 +a^2_{Z,\cross}z_{2}z_1]\Big(
n_H\bar{\beta}_2\partial_{h_2}f+\frac{(\bar{\omega}_2)^2}{2}\partial^2_{z_2z_2}f\Big)
\end{align*}
One can also include co-jumps, which would change the coefficients above and add a term of the form $\rho \partial^2_{z_1z_2}f$, where $\rho$ is the correlation between the Poisson processes driving $1$ and $2$.

When considering only ZHawkes without Hawkes ($n_H=0$), we have $h_1=h_2=0$, thus the infinitesimal generator results in:
\begin{align*}
    Af^\infty(z_{1},z_{2}) =&
 -\bar{\omega}_{1}z_{1}\partial_{z_{1}}f 
 -\bar{\omega}_{2}z_{2}\partial_{z_{2}}f \\
&+ [\lambda^1_\infty  + (a^1_{Z,1}z_1)^2 + (a^1_{Z,2}z_2)^2 +a^1_{Z,\cross}z_{2}z_1]
\frac{(\bar{\omega}_1)^2}{2}\partial^2_{z_1z_1}f
+ [\lambda^2_\infty + (a^2_{Z,1}z_1)^2 + (a^2_{Z,2}z_2)^2 +a^2_{Z,\cross}z_{2}z_1]
\frac{(\bar{\omega}_2)^2}{2}\partial^2_{z_2z_2}f
\end{align*}



\subsection{General ODE for $F(\theta)$}\label{app:ODE_F}

We write here the general ODE on $F(\theta)$ for any parameters $a_Z$'s and $\omega$'s:

\begin{equation}\label{eqApp:ODEonF}
    \begin{aligned}
&\Big(\omega_1+ \omega_2 + (a_{Z,1}^1 \omega_1)^2   + (a_{Z,2}^2 \omega_2)^2 - \alpha
    (\omega_1 +2 (a_{Z,1}^1\omega_1)^2 )  \cos^2(\theta)  - \alpha(\omega_2+2(a_{Z,2}^2\omega_2)^2) \sin^2(\theta) 
     \\& - (a_{Z,\cross}^2 (\omega_2)^2+a_{Z,\cross}^1 (\omega_1)^2 )\alpha\cos\theta  \sin\theta 
      \\&+
    \frac{1}{8} \alpha(\alpha+(2+\alpha)\cos(2\theta))(\omega_1)^2\left((a_{Z,1}^1)^2 + (a_{Z,2}^1)^2 + ((a_{Z,1}^1)^2 - (a_{Z,2}^1)^2) \cos(2 \theta) + 
    a_{Z,\cross}^1  \sin(2 \theta)\right)\\& + 
    \frac{1}{8} \alpha(\alpha-(2+\alpha)\cos(2\theta))(\omega_2)^2 \left((a_{Z,1}^2)^2 + (a_{Z,2}^2)^2 + ((a_{Z,1}^2)^2 - (a_{Z,2}^2)^2) \cos(2 \theta) + 
    a_{Z,\cross}^2  \sin(2 \theta)\right)\Big) F(\theta)\\& + \Big(a_{Z,\cross}^2  (\omega_2)^2 \cos^2(\theta) - a_{Z,\cross}^1  (\omega_1)^2 \sin^2(\theta)  - 
    \cos\theta  \sin\theta \left(\omega_1  - \omega_2 + 
    2 (a_{Z,1}^1 \omega_1)^2 - 2 (a_{Z,2}^2 \omega_2)^2 \right) 
    \\& + 
    \frac{1}{4} (1 + \alpha) (\omega_1)^2 \sin(2 \theta) ((a_{Z,1}^1)^2 + (a_{Z,2}^1)^2 + ((a_{Z,1}^1)^2 - (a_{Z,2}^1)^2 ) \cos(2 \theta) + 
       a_{Z,\cross}^1  \sin(2 \theta)) \\&- 
    \frac{1}{4} (1 + \alpha) (\omega_2)^2 \sin(2\theta) ((a_{Z,1}^2)^2 + (a_{Z,2}^2)^2 + ((a_{Z,1}^2)^2 - (a_{Z,2}^2)^2) \cos(2 \theta) + 
       a_{Z,\cross}^2  \sin(2 \theta))\Big) F'(\theta) \\&+ 
 \frac{1}{4} \Big((\omega_1)^2 \sin^2(\theta) ((a_{Z,1}^1)^2 + (a_{Z,2}^1)^2 + ((a_{Z,1}^1)^2 - (a_{Z,2}^1)^2)  \cos(2 \theta) + 
       a_{Z,\cross}^1  \sin(2 \theta)) \\&+ 
    (\omega_2)^2 \cos^2(
      \theta) ((a_{Z,1}^2)^2 + (a_{Z,2}^2)^2 + ((a_{Z,1}^2)^2 - (a_{Z,2}^2)^2)  \cos(2 \theta) + 
       a_{Z,\cross}^2  \sin(2 \theta))\Big) F''(\theta)=0
    \end{aligned}
\end{equation}

\end{document}

%% file: preamble.tex
\usepackage{amsthm}
\usepackage{mathtools}
\usepackage{physics}
\usepackage{xcolor}
\usepackage{graphicx}
\usepackage[left=23mm,right=13mm,top=35mm,columnsep=15pt]{geometry} 
\usepackage{adjustbox}
\usepackage{placeins}
\usepackage[T1]{fontenc}
\usepackage{lipsum}
\usepackage{csquotes}